\documentclass[10pt]{iopart}

\usepackage[switch]{lineno}

\usepackage{graphicx}
\usepackage[utf8]{inputenc}
\usepackage{textcomp}

\usepackage{booktabs}
\usepackage{rotating}
\usepackage{multirow}
\usepackage{subfigure}

\usepackage[sort&compress,comma,square,numbers]{natbib}                                
\usepackage{natbib,url,hyperref}
\hypersetup{
     colorlinks   = true,
     citecolor    = blue,
     linkcolor    = blue,
     urlcolor     = blue
}

\usepackage{color}
\usepackage[usenames,dvipsnames,svgnames,table]{xcolor}

\begin{document}

\frenchspacing

\title{Tungsten migration energy barriers for surface diffusion: a parameterization for KMC simulations}

\author{Ville Jansson$^1$, Andreas Kyritsakis$^1$, Simon Vigonski$^{1,2}$, Ekaterina Baibuz$^1$, Vahur Zadin$^2$, Alvo Aabloo$^2$, and Flyura Djurabekova$^1$}

\address{$^1$ Helsinki Institute of Physics and Department of Physics, P.O. Box 43\\ (Pehr Kalms gata 2), FI-00014 University of Helsinki, Finland}

\address{$^2$ Institute of Technology, University of Tartu, Nooruse 1, 50411 Tartu, Estonia}

\ead{ville.b.c.jansson@gmail.com}

\begin{abstract}
We have calculated the migration barriers for surface diffusion on Tungsten. 
Our results form a self-sufficient parameterization for Kinetic Monte Carlo simulations of arbitrarily rough atomic tungsten surfaces, as well as nanostructures such as nanotips and nanoclusters.
The parameterization includes first- and second-nearest neighbour atom jump processes, as well as a third-nearest neighbour exchange process. 
The migration energy barriers of all processes are calculated with the Nudged Elastic Band method.
The same attempt frequency for all processes is found sufficient and the value is fitted to Molecular Dynamics simulations.
The model is validated by correctly simulating with Kinetic Monte Carlo the energetically favourable W nanocluster shapes, in good agreement with Molecular Dynamics simulations.

\vspace{1cm}

\end{abstract}

%\submitto{\NT}
\maketitle
\ioptwocol	% For two column layout
% \linenumbers

\section{Introduction}

% Tugsten nanofeatures and surfaces are important 
Tungsten is one of the most commonly used materials in various applications of vacuum electronics, spanning from electron and ion sources to atom probes and X-ray tubes, mainly due to its hardness, high density, high melting temperature and resistance to oxidation.
Due to the aforementioned importance, the evolution of tungsten surfaces due to self-diffusion has attracted the interest of many researchers for many decades \cite{boling1958blunting,barbour1960determination, tsong1975direct,wang1982field,yeong2006field,fujita2007mechanism,suzuki2018single,vurpillot2018simulation}.

% Kimocs model and the usefulness of a parameterization
Although theoretical diffusion models have been used to describe the macroscopic evolution of W surfaces \cite{barbour1960determination}, an accurate description of the underlying processes at the atomic level is necessary to understand the evolution of nanoscale structures \cite{vigonski2018au}.
Such a description can be achieved by means of atomistic simulations such as Density Functional Theory (DFT), Molecular Dynamics (MD) and Kinetic Monte Carlo (KMC).

Kimocs \cite{jansson2016long,kimocs} is an open-source KMC code that has been shown to be an effective way to study the thermal evolution, driven by atom diffusion, of atomic surfaces and nanostructures.
Kimocs has been used to simulate Cu nanotips and nanowires \cite{jansson2016long,baibuz2018migration,lahtinen2018artificial}, Au nanowires \cite{vigonski2018au}, and Fe nanoclusters \cite{zhao2016formation}; showing good agreements with both Molecular Dynamics (MD) simulations and experiments.

In order to simulate any atomic system with Kimocs, a parameterization containing the migration energy barriers and attempt frequencies for all the considered atom transitions, is needed.
The parameterization gives the physical properties of the simulated material and thus plays a similar role for KMC as the interatomic potentials plays in MD simulations.
Several such pre-calculated parameter sets for different materials have previously been developed and used in other Kimocs studies \cite{jansson2016long,baibuz2018migration,zhao2016formation, vigonski2018au}. 
For other parameterization approaches in atomistic KMC, see the references in \cite{jansson2016long}.  

The attempt frequencies are commonly taken to be the same for all atom jumps in KMC models, as was done in \cite{jansson2016long,baibuz2018migration, vigonski2018au, zhao2016formation}.
It has however been suggested that the attempt frequencies may be orders of magnitudes higher for atom jumps longer than the first-nearest neighbour distance \cite{antczak2004long}.
This result by G\;Antczak and G\;Ehrlich was obtained by multi-variable fitting of KMC simulations results to Field Ion Microscopy data of W adatom diffusion on the \{110\} surface; long jumps on the \{100\} surface, where only long jumps are possible, were not studied.

The migration energy barriers may be calculated using MD and the Nudged Elastic Band method \cite{mills1994quantum,mills1995reversible}.
Kimocs uses a rigid atomic lattice and the transitions are defined by counting the number of neighbour atoms (see section \ref{sec:methods}). 
With this parameterization, between 1000 and 7000 first-nearest neighbour jump barriers need to be calculated for a complete parameterization. 
For the body-centred cubic (bcc) lattice, such as for Fe and W, also longer transitions, such as the second-nearest neighbour jumps may play a important role. 
In \cite{zhao2016formation, baibuz2018migration}, a small number of second-nearest jumps were included in the simulations of the bcc Fe nanoclusters, along the first-nearest neighbour jumps, as the bcc\{100\} surfaces were important in the study.
Even though the number of second-nearest neighbour barriers were quite small, the simulations still showed good agreement with the experiments \cite{zhao2016formation, baibuz2018migration}.

% Research statement (filling the gap)
In this work we present a self-sufficient KMC parameterization of the atom transitions on tungsten surfaces, which allows KMC simulations of arbitrary tungsten surfaces and other nanostructures, such as nanoclusters.
The parameterization includes large sets of calculated first- and second-nearest neighbour atom jump barriers, as well as a third-nearest neighbour exchange process that is likely to be important on \{100\} surfaces.
We will show that using one attempt frequency for all atom transitions gives sufficient precision and we will fit this attempt frequency by comparing to MD simulations of nanotips (sections \ref{sec:attempt_long} and \ref{sec:attempt}). 
We will validate the parameterization by showing that it predicts correctly in KMC simulations the energy minimum shapes of W nanoclusters (section \ref{sec:wulff}).
Finally, we will discuss the results (section \ref{sec:discussion}) and summarize our conclusions (section \ref{sec:conclusions}).

\section{Methods}\label{sec:methods}

In this work we used the open-source KMC code Kimocs \cite{jansson2016long,kimocs}. 
Kimocs was specially developed to study atom diffusion processes on metallic surfaces. 
It uses a rigid lattice for the atoms, that are either face-centred-cubic (fcc) or body-centred-cubic (bcc).
Atom transitions up to third-nearest neighbour distance vacant lattice sites, as jumps or exchange processes, may currently be considered by the code.
In this paper we use a bcc lattice and the considered atom transitions, as included in the W parameterization, are described in detail in section \ref{sec:parameterization}. 

The probability rate of any atom transition event in KMC is calculated using Arrhenius' equation
\begin{equation}\label{eq:arrhenius}
\Gamma = \nu\exp \left(\frac{-E_m}{k_B T}\right),
\end{equation}
where $T$ is the temperature of the system, $k_B$ is the Boltzmann constant, $E_m$ is the migration energy barrier, and $\nu$ the attempt frequency. 
The migration energy barriers $E_m$ are different for all atom transition processes and need to be tabulated prior to the KMC simulations.
The individual atom transitions are described in the Kimocs parameterization by counting the number of first and second nearest neighbours of the initial position ($a$ and $b$) as well as the final position ($c$ and $d$). 
This parameterization is also called the 4d parameterization for short \cite{baibuz2018migration}.
We only consider pure metals in our simulations. 

\subsection{Molecular Dynamics and Nudged Elastic Band calculations}\label{sec:neb}
All MD simulations were done using LAMMPS \cite{lammps}.
The set of migration barriers $E_m$ for W was calculated with LAMMPS using the Nudged Elastic Band (NEB) method \cite{mills1994quantum,mills1995reversible} with the tethering approach \cite{baibuz2018migration}, where the barriers are calculated on an atomic lattice which has been made semi-rigid by tethering every atom in the system with a small tethering force.
The tethering force is not explicitly included in the calculation of the system potential energy nor the energy barrier.
This approach avoids the discrepancy between describing processes in unstable atomic configurations, where atoms may relax to other lattice positions, and in a rigid lattice, such as the one used in Kimocs, while still keeping the systematic error, which may be introduced by the extra tethering forces, at a minimum. 
The tethering method is extensively described and discussed in \cite{baibuz2018migration} and has also been successfully used before to reproduce experimental results for Au nanowires \cite{vigonski2018au}.

The first-nearest neighbour atom jump processes were calculated with NEB on a substrate with either a \{110\}, \{100\}, or a \{111\} surface, or in a bulk system, where neighbour atoms have been added or removed to give all possible $(a,b,c,d)$ configurations.
All first-nearest neighbour jumps on other surfaces will be covered by the ``bulk'' calculations, where the atom, even if it is really inside a void, can be set on an arbitrary surfaces by removing neighbour atoms.
The second-nearest neighbour jump processes were calculated using the bulk method, whereas the third-nearest exchange process was calculated on a \{100\} surface.
If several processes with the same jump distance can be described by the same $(a,b,c,d)$ label, the one with the lowest barrier value is used; barriers calculated on one of the \{110\}, \{100\} or \{111\} surface substrates are however prioritized to barriers calculated in the bulk \cite{baibuz2018migration}.

For the NEB barrier calculations and all MD simulations, we used the EAM4 W potential from \cite{marinica2013interatomic}. 
The implementation of the potential also included a short range stiffening from \cite{sand2016non}, but that would only affect atoms with kinetic energies above 10\;eV, which would not appear in our 0\;K NEB calculations. 
We also tried the interatomic potential by Derlet et al. \cite{derlet2007multiscale}, but the convergence of the NEB calculations with this potential was found to be too slow for this work, where we need to calculate thousands of barriers.

The calculated parameter set for W is described in section \ref{sec:parameterization}.

\section{Results}\label{sec:results}

\subsection{W parameterization}\label{sec:parameterization} 
% R20181115_marinica
 
Our W parameterization includes all migration energy barriers and their attempt frequencies, for all atom transitions needed to correctly simulate W surfaces with KMC.
As discussed in section \ref{sec:attempt}, we will use the same attempt frequency for all transitions.
The parameterization includes three separate subsets of barriers: first-nearest neighbour jumps, second-nearest neighbour jumps, and a third subset with only a single, but still important, third-nearest neighbour exchange process.

\begin{figure}
 \centering
 \includegraphics[width=\columnwidth]{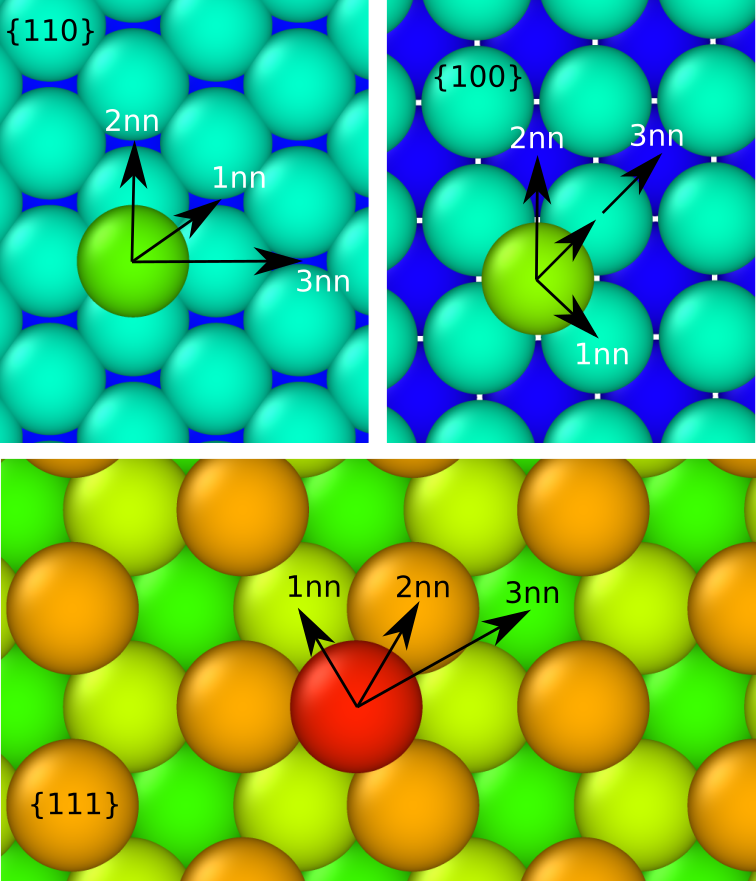}\label{fig:long_transitions}
\caption{Examples of atom transitions on the \{110\} surface (left), the \{100\} surface (right), and the \{111\} surface (below). 
On the \{110\} surface, the first-nearest neighbour (1nn) jump is the most common jump and is in our parameterization labelled as $(2,2,3,2)$.
The 2nn and 3nn jumps would be labelled $(2,2,2,3)$ and $(2,2,2,2)$, respectively, although the latter jump is not included in our parameter set.
On the \{100\} surface (right), the 1nn jumps would be labelled $(4,1,1,1)$ and end up in an unstable position with only a single second-nearest neighbour, which makes the jump rather improbable.
More likely on this surface is the second-nearest (2nn) $(4,1,4,2)$ jump.
A third possible transition is that the atom may replace one of the (light green) atoms in the lower layer, which is pushed diagonally in a concerted exchange process to a third-nearest neighbour position (3nn).
In this work, we consider such an exchange process, involving two atoms, as a single 3nn process with a single migration energy, where the actual exchange will be ignored. 
This 3nn exchange process has the label $(4,1,4,1$. An animation of this process from the NEB calculation is included in the Supplementary Material.
The \{111\} surface (lower panel) is the least energetically favourable of the low-index surfaces (see table \ref{table:surface_energies}). 
On the lower panel are shown the 1nn $(4,3,2,3)$, 2nn $(4,3,1,2)$, and 3nn $(4,3,4,3)$ jump processes of the red atom, although only the first two are possible in our current model.
}
\label{fig:processes}
\end{figure}

\begin{figure}
 \centering
 \includegraphics[width=\columnwidth]{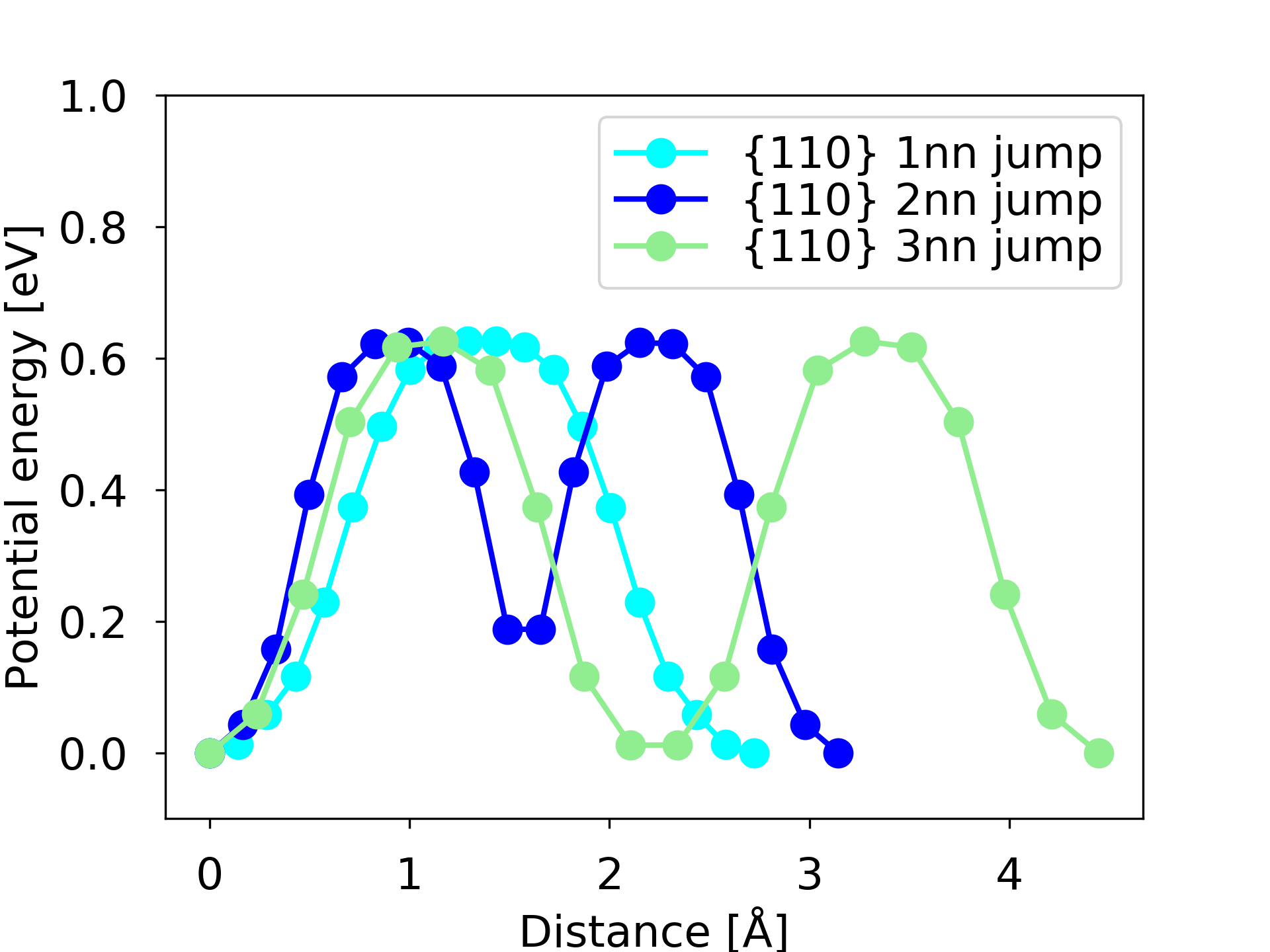}
 \caption{Migration energy barriers, calculated with NEB, for the first- (1nn), second- (2nn), and third-nearest neighbour (3nn) atom jumps on the \{110\} surface.
 The 2nn and 3nn jump path show an energy minimum in the centre as the atom actually makes two subsecquent 1nn jumps via a bcc site in order to finish the transition, which is also why the barriers have the same height.
 The $x$-values are projected to the linear distance between the initial and final position of the jumps, whereas the actual path of the 2nn and the 3nn jumps are in reaity longer.}
 \label{fig:unlikely_processes}
\end{figure}

In figure \ref{fig:processes} are shown example transitions on the low-index surfaces \{110\}, \{100\} and \{111\}.
On the \{110\} surface (left panel), the first-nearest neighbour atom jump will dominate due to its low migration energy barrier 0.63\;eV.
The second and third-nearest neighbour transitions are also marked for reference, but in our NEB calculations we see that these transitions have a minimum energy path that goes as two subsequent first-nearest neighbour atom jumps.
The migration energy barriers for these long jumps are compared to the first-nearest neighbour atom jump in figure \ref{fig:unlikely_processes}.

On the plain \{100\} surface (right panel in figure \ref{fig:processes}), an adatom may only make a first-nearest jump to an unstable position with only one nearest neighbour. Therefore, second-nearest neighbour jumps and the third-nearest neighbour exchange process are more important on this surface.
In principle, second-nearest neighbour exchange processes are also possible, but we assume them to have generally higher energy barriers (making them less likely) than the corresponding jump processes, as is the case for the plain surface case (2.73\;eV versus 2.64\;eV), and ignore them in this model.
The diagonal third-nearest neighbour exchange process has a migration energy barrier of 1.95\;eV, which is even lower than the second-nearest neighbour jump and thus important for the \{100\} surface.

To get a full physical description of the diffusion-driven evolution of a W surface, all three subsets need to be included in the used parameterization in the KMC simulations.
The three barrier subsets are included in the Supplementary Materials.

\subsubsection{First-nearest neighbour jumps}
% R20190408-neb  

\begin{figure*}
  \centering
\subfigure[First-nearest neighbour jumps]{
  \includegraphics[width=0.45\textwidth]{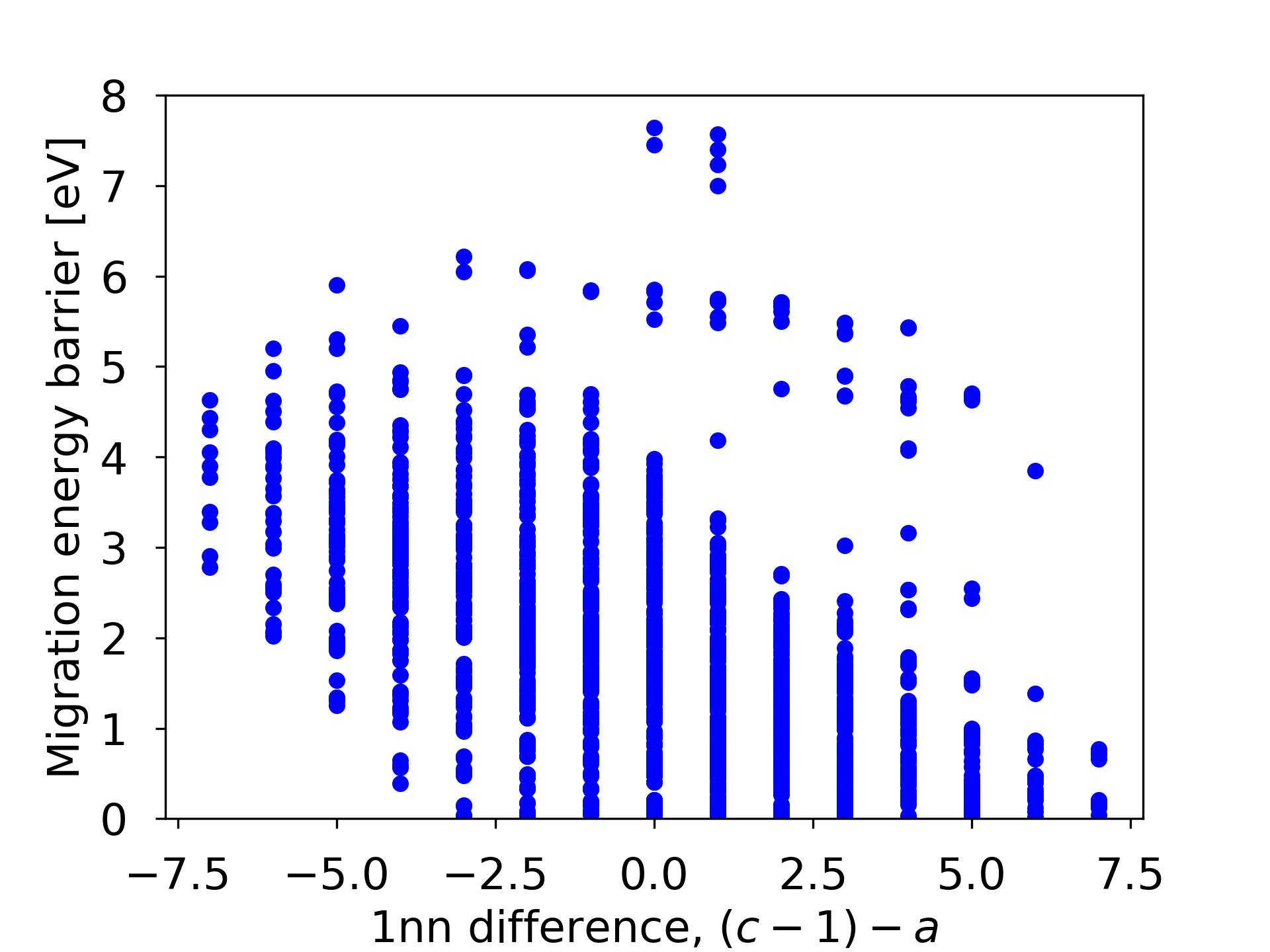}\label{fig:1nn}
}
\subfigure[Second-nearest neighbour jumps]{
  \includegraphics[width=0.45\textwidth]{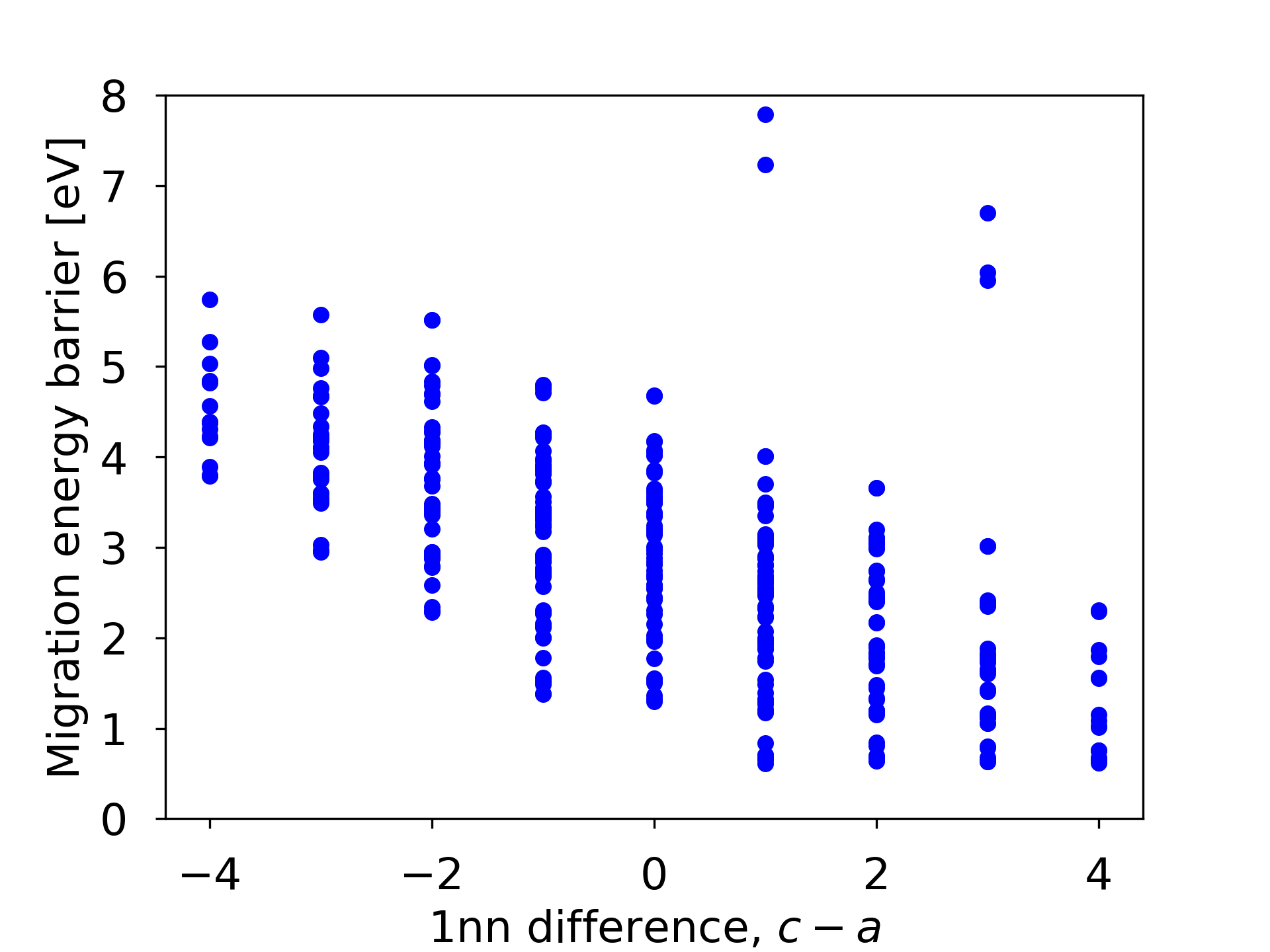}\label{fig:2nn}
}

\caption{The migration energy barriers versus the difference of number of first-nearest neighbours (1nn) of the jumping atom, after and before the jump (not counting the jumping atom itself as a neighbour for the 1nn jumps). The first-nearest neighbour jumps (1nn) are shown in (a) and the second-nearest neighbour jumps (2nn) in (b).}
\label{fig:barriers}
\end{figure*}

For the first-nearest neighbour distance jump processes, 1734 migration energy barriers were calculated with NEB and are shown in figure \ref{fig:1nn}.
The different jumps are identified by the number of first- ($a$ and $c$) and second-nearest ($b$ and $d$) neighbour atoms of the initial atom lattice position and the target vacancy position, giving four integer values describing every process, $(a,b,c,d)$.
The jumping atom will thus be counted as a first-neighbour atom of the target vacancy ($c$).
Jumps to a target position with no first-nearest neighbours ($c \leq 1$) are not included, as we do not consider evaporation processes in this work.
If an atom would somehow end up in a position with no first- or second-nearset neighbours, it will not be able to do any jump.
For the current parameterizaton, this almost never happens, but as a default Kimocs would automatically remove such artifact atoms from the system.
The $(a,b,c,d)$ parameters takes values between 
$0 \leq a \leq 7$, 
$1 \leq b \leq 6$, 
$2 \leq c \leq 8$, and 
$0 \leq d \leq 6$.
A process with the maximum number of neighbours $(7,6,8,6)$ would be equvalent to a vacancy jump in complete bulk. 

\subsubsection{Second-nearest neighbour jumps}
% R20190501_2nn

Second-nearest neighbour distance jump processes are particularly important on the bcc\{100\} surface, as no first-nearest neighbour jumps are possible on this surface, but these long processes may also play a role on other surfaces and e.g. on edges between different facets.
For this subset, 391 second-nearest neighbour jump barriers were calculated, which includes all possible surface jumps, but not clear-cut bulk processes.
The barriers are shown in figure \ref{fig:2nn}.
The jump processes are again identified by the $(a,b,c,d)$ parameters; the same way as the first-nearest neighbour jumps, although here the jumping atom will of course be counted as one of the second-nearest neighbour atoms of the target vacancy ($d$).
Processes with two or less first-nearest neighbour atoms ($a,c \leq 2$) of the initial or target vacancy position were not included as these positions are not likely to be stable. 
Jumps from and to such unstable positions are however already allowed via first-nearest neigbour jumps.
Of the first-nearest neighbour atom positions of the initial and final position, four positions are common and their atoms are counted in both $a$ and $c$.
We only considered cases with two such common neighbours, as less would correspond to a non-continuous surface, and more than two common atoms would correspond to a bulk case where a second-nearest neighbour atom jump would likely be blocked by these neighbour atoms from happening in the first place.
Cases with ($b > 5$) where also discarded, as this is only possible for bulk processes and thus beyond the scope of this work.
All in all, the $(a,b,c,d)$ parameters takes values between 
$2 \leq a \leq 6$, 
$1 \leq b \leq 4$, 
$2 \leq c \leq 6$, and 
$2 \leq d \leq 5$.

\subsubsection{Third-nearest neighbour exchange processes}

The diagonal exchange process (3nn in figure \ref{fig:processes}), where two atoms move in concert so that the effect is that one atom appears to have made a third-nearest neighbour distance jump, may be important on the bcc\{100\} surface and is therefore also included in the parameterization.
The exchange process is described the same way as the first- and second-nearest jumps, by counting the neighbours: $(4,1,4,1)$. 
The barrier was calculated with NEB to be 1.96 eV, which is less than the 2.64 eV barrier of the simplest bcc\{100\} second-nearest neighbour atom jump $(4,1,4,2)$.
Calculating other exchange processes or third-nearest neighbour processes would of course further increase the precision of the model, but this is beyond the scope of this work.

In the KMC simulations with Kimocs, an exchange process will be treated as a single-atom jump with the difference that the barrier is the one calculated with NEB for the actual process, where the exchange was taken into account.
The actual exchange of atoms in Kimocs is ignored, but as all atoms are of the same element, this will not change the evolution.

\subsubsection{Attempt frequencies of long vs short transitions}\label{sec:attempt_long}

Additionally to the migration energy barrier $E_m$, the attempt frequency $\nu$ is needed to give the correct probability rate for any atom transition, as described by equation \ref{eq:arrhenius}.
For KMC models, where only first-nearest neighbour jumps have been included, it has been found sufficient to only use a single $\nu$ value for all jumps.
However, in our parameterization we are also using second-nearest neighbour jumps, as well as a third-nearest neighbour exchange process, which raises the question whether it is still a good approximation to only use a single $\nu$ value.

G\;Antczak and G\;Ehrlich \cite{antczak2004long} fitted attempt frequencies of atom long jumps on a W\{110\} surface by comparing KMC simulations with Field Ion Microscopy diffusion results.
Their fitted attempt frequency for second-nearest neighbour atom jumps was four orders of magnitudes larger than for first-nearest jumps; for a third-nearest jump, they found the attempt frequency to be seven orders of magnitudes larger.

\begin{figure}
 \centering
 \includegraphics[width=\columnwidth]{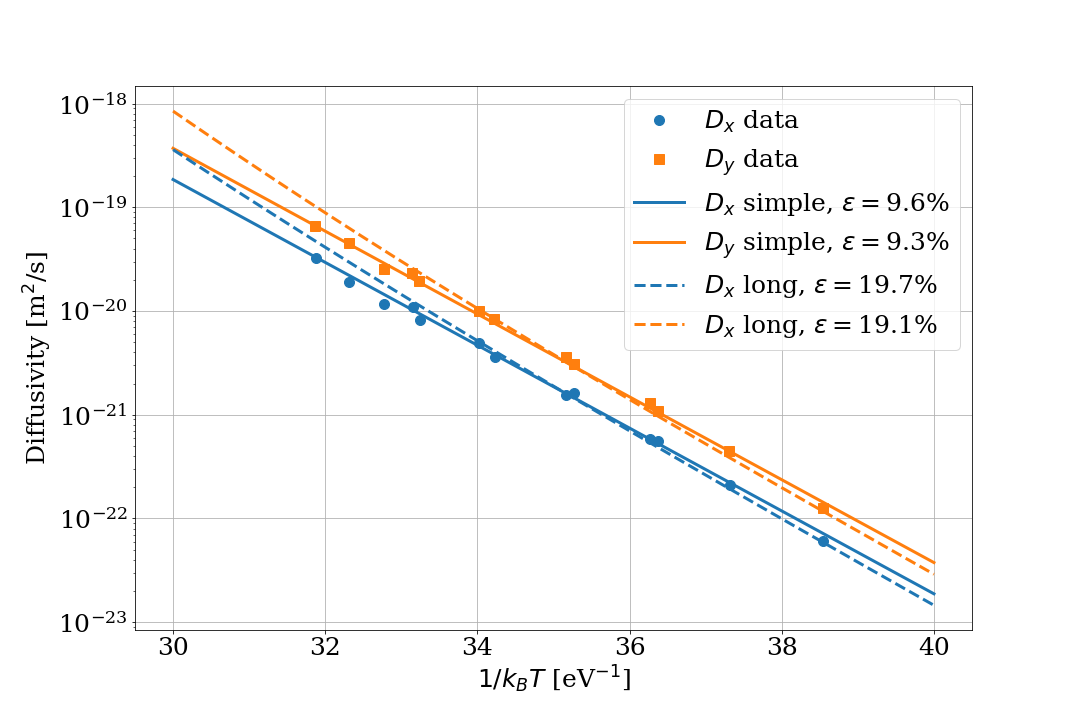}
 \caption{Diffusivity data along x and y directions as reported in \cite{antczak2004long} (markers), fitted with two different models: the simple-jump model which considers only first-nearest neighbour jumps (solid lines) and the long-jump model considered in \cite{antczak2004long} (dashed lines).}
 \label{fig:diffusivities}
\end{figure}

However, their analysis significantly overestimates the contribution of the long jumps in the diffusion process.
In figure \ref{fig:diffusivities} we plot the diffusivity data reported in figure 1 in \cite{antczak2004long} (markers).
The dashed lines represent the long-jump model in \cite{antczak2004long}, which includes jumps on the W\{110 \} surface (denoted as $\beta,\delta_x$ and $\delta_y$ jumps), with the diffusivities being calculated according to equation (1,2) in \cite{antczak2004long}, using the barriers and attempt frequencies extracted in \cite{antczak2004long} for each jump.
The solid lines represent a simple-jump model which includes only first-nearest neighbour jumps (denoted as $\alpha$ in \cite{antczak2004long}) with a fitted barrier of 0.92 eV and an attempt frequency $\nu = 3.52 \times 10^{12}$\;s$^{-1}$.

The relative error of each curve is shown in the legend.
We see that the simple model reproduces the experimental data with about half the error, as compared to the long-jump model.
Furthermore, the line of the long-jump model is curved upwards, which is not expected normally for the diffusivity on a metal surface.
In fact, the long-jump model predicts a diffusivity of the order of $10^{-3}$\;m$^2/s$ at $T=3000$\;K, which is not physical for a solid metal.
Therefore, we may conclude that the attempt frequencies of the order of $10^{16}$ -- $10^{19}$\;s$^{-1}$ extracted in \cite{antczak2004long} are significantly overestimated.

A method to calculate the attempt frequencies is to simulate the diffusion of a single adatom defect on a perfectly flat surface.
Since we already calculated the (non-tethered) migration energy barriers $E_m$ for the adatom transitions listed in table \ref{table:barriers}, we can count the number of jumps $N_j$ of a particular kind that the adatom does during a time period $t$. 
We can then calculate the probability rate for this kind of jump as
\begin{equation}\label{eq:rate}
 \Gamma = \frac{N_j}{tN_p},
\end{equation}
where $N_p$ is the number of different directions that the adatom can do the jump at any lattice point.
For a perfect W\{110\} or W{100\} surface, with only one adatom, we will have $N_p = 4$ for any jump of the adatom. 
The attempt frequency of the jump can then be obtained as
\begin{equation}
 \nu = \Gamma\exp\left(\frac{E_m}{k_BT}\right).
\end{equation}
\begin{table*}
  \centering
   % R20181115_marinica_no_tethering
   % R20190819 (MD)
   \caption{Migration energy barriers for three different processes, calculated with NEB, with and without the use of tethering, and compared to DFT values.
   The DFT values in parenthesis are calculated for a ``zigzag'' reconstructed \{100\} surface, which may occur below 250\;K \cite{debe1977space,heinola2010first,olewicz2014coexistence} (see discussion in section \ref{sec:discussion}). 
   The attempt frequencies are estimated from MD diffusion simulations and by the harmonic approximation (HA) of the non-tethered data.
   The errorbars are the standard deviations of six MD runs with different seeds.}
   \label{table:barriers}
   \begin{tabular*}{\textwidth}{@{\extracolsep{\fill}} l c c c c c c}
   \toprule
   Process & Surface & $E_m$ & $E_m$ (Tet.) & $E_m$ (DFT)                               & $\nu$ (MD) & $\nu$ (HA)\\
           &         & [eV]       & [eV]    & [eV] & [s$^{-1}$] & [s$^{-1}$]\\
   \midrule
    (2,2,3,2) 1nn jump     & \{110\} & 0.63 & 0.63 & 0.87$^a$                           & $(1.1\pm0.1)\cdot10^{12}$& $1.5\cdot10^{12}$\\
    (4,1,4,2) 2nn jump     & \{100\} & 2.64 & 2.66 & 2.27$^a$, 2.47$^{b}$, (2.49$^{b}$) & $(2.7\pm2.2)\cdot10^{13}$& $1.9\cdot10^{12}$ \\
    (4,1,4,1) 3nn exchange & \{100\} & 1.95 & 1.96 & 1.55$^{b}$, (2.05$^{b}$)           & $(8.5\pm2.2)\cdot10^{12}$& \\
   \bottomrule
   \end{tabular*}
   \flushleft{${}^a$ \cite{chen2013biaxial}, ${}^{b}$ \cite{olewicz2014coexistence}. }
\end{table*}

Our atomic system consisted of a single W adatom on square W\{110\} or W\{100\} surface with both sides 5\;nm and a thickness of 2\;nm.
The lowest layer of atoms were fixed.
The system was simulated with constant volume and temperature at 2500\;K. 
The lattice parameter of the atomic system was obtained in a separate simulation by relaxing the volume of a bulk system at constant zero pressure at 2500\;K.

For the \{110\} case, the adatom was allowed to diffuse for 200\;ps before the simulation was stopped (figure \ref{fig:md_diffusion}).
This was repeated 6 times with different seeds.
For the \{100\} case, the diffusion was dominated by third-nearest neighbour exchange processes with only a few second-nearest neighbour jumps; first-nearest neighbour jumps not being possible at all on this particular surface.
The adatom defect (which changed atom after every exchange process) was traced in Ovito \cite{ovito} and the number of (4,1,4,1) and (4,1,4,2) transitions were counted.
Other adatoms and vacancies were occasionally created due to the high temperature.
The transitions of the adatom defect were only counted until the adatom defect was blocked when another adatom came within a third-nearest neighbour distance and thus blocking the jump possibilities of the diffusing adatom defect.

The average jump rates $\Gamma$ for the different simulations are listed in table \ref{table:gamma} and the obtained attempt frequencies $\nu$ are listed in table \ref{table:barriers}.
The attempt frequencies are found to differ slightly more from each other with the largest difference of a factor $\sim$25 found between the first- and the second-nearest jumps.
\begin{figure}
 \centering
 \includegraphics[width=\columnwidth]{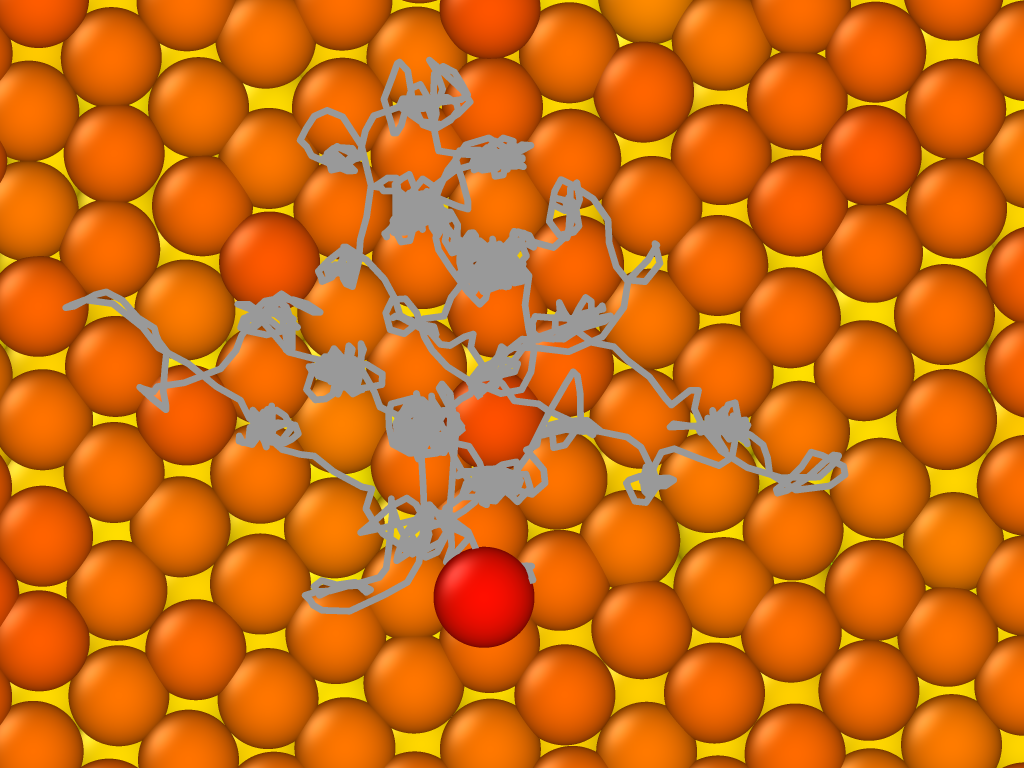}
 \caption{MD simulation of an adatom diffusing on a \{110\} surface for 200\;ps at $T=2500$\;K.
 The atoms are coloured according to their $z$ coordinate, except for the diffusing atom, which is coloured red. 
 The trace of the diffusing atom is shown as a grey line.
 The figure is made with Ovito \cite{ovito}.}
 \label{fig:md_diffusion}
\end{figure}
\begin{table*}
  \centering
   % R20190819
   \caption{The average probability rates $\langle\Gamma\rangle$, as defined in equation \ref{eq:rate}, calculated for different processes from MD simulations of a single diffusing adatom defect on a perfect surface.
   The average times $\langle t \rangle$ for the considered adatom defect diffusion paths are also shown.
   The \{100\} data are from the same set of simulations.
   The errors are estimated as the standard deviation of 6 separate simulations of each surface case, using different seeds.}
   \label{table:gamma}
\begin{tabular*}{\textwidth}{@{\extracolsep{\fill}} l c c c}
\toprule
Process & Surface & $\langle\Gamma\rangle$ & $\langle t \rangle$\\
        &         & [s$^{-1}$] & [s]\\
\midrule
    (2,2,3,2) 1nn jump     & \{110\} & $(6.0\pm0.6)\cdot10^{10}$ & $(2.0\pm1.6)\cdot10^{-10}$\\
    (4,1,4,2) 2nn jump     & \{100\} & $(1.3\pm1.0)\cdot10^8$    & $(3.5\pm1.6)\cdot10^{-9}$\\ 
    (4,1,4,1) 3nn exchange & \{100\} & $(1.0\pm0.3)\cdot10^9$    & $(3.5\pm1.6)\cdot10^{-9}$\\
\bottomrule
\end{tabular*}
\end{table*}

A second way to roughly estimate the attempt frequency is by using the harmonic approximation method where the adatom at the initial equilibrium lattice position is approximated as a harmonic oscillator with a potential
\begin{equation}\label{eq:harmonic}
 U = \frac{1}{2}kx^2,
\end{equation}
where $k$ is a spring constant and $x$ the displacement from equilibrium.
By fitting equation \ref{eq:harmonic} to the NEB calculated potential landscape of the one-dimensional atom transition, as seen in figure \ref{fig:neb_barriers}, $k$ can be obtained and the attempt frequency $\nu$ can be estimated as \cite{liu1991eam}
\begin{equation}
 \nu = \frac{1}{2\pi}\sqrt{\frac{k}{m}},
\end{equation}
where $m$ is the mass of the W atom.
In figure \ref{fig:neb_barriers} are shown the NEB-calculated minimum energy paths (no tethering is used) for the first-nearest neighbour (2,2,3,2) atom jump and the second-nearest neighbour (4,1,4,2) atom jump.
The harmonic fits are shown as black curves around the initial position.
The resulting attempt frequencies are presented in table \ref{table:barriers}.
The values for the jump processes are found to be between $1.5\cdot10^{12}$ and $1.9\cdot10^{12}$ s$^{-1}$ and thus within the same order of magnitude, but slightly lower as found previously with the MD diffusion simulations.
\begin{figure}
 \includegraphics[width=\columnwidth]{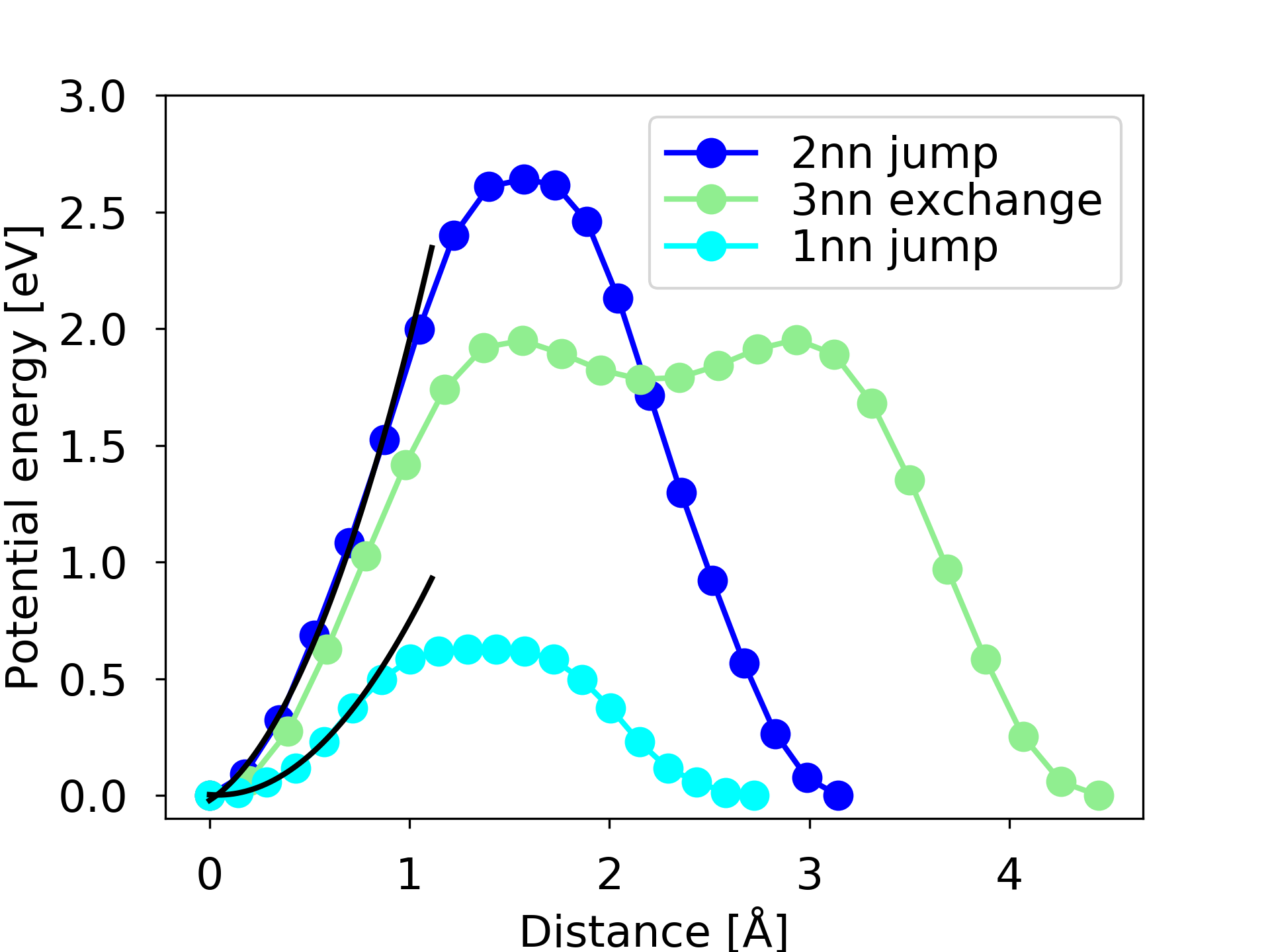}
 \caption{The migration energy barriers for the first-nearest neighbour (1nn) distance (2,2,3,2) jump process on the \{110\} surface, the second-nearest neighbour (2nn) distance (4,1,4,2) jump process on the \{100\} surface, and the third-nearest neighbour (3nn) distance (4,1,4,1) exchange process on the \{100\} surface.
 The black lines are the harmonic fits. 
 These barriers were calculated with NEB without tethering.}
 \label{fig:neb_barriers}
\end{figure}

\subsubsection{Fitting of a global attempt frequency}\label{sec:attempt}

% R20190522-nu_model5
% R20190401-nu

Since calculating the attempt frequencies for all thousands of atom transitions included in our model is rather unfeasible, we will instead fit a global attempt frequency $\nu$ which we will use for all processes, similarly as have been done in previous KMC models \cite{jansson2016long,baibuz2018migration,vigonski2018au}. 
The error from doing this will be minimal as we previously established that the attempt frequencies of long jumps and exchange processes are not expected to differ much more than an order of magnitude from each other.
This global $\nu$ will only affect the time estimates of the KMC model and is thus merely a scaling factor and only roughly correspond to the attempt frequencies calculated for individual processes (table \ref{table:barriers}). 
This frequency $\nu$ was obtained by simulating the flattening process of a nanotip [figure \ref{fig:cylinder_a}] with both MD and KMC and fitting the KMC attempt frequency so that both simulation techniques gave the same flattening time value.
This way, the time estimate of the KMC model can be expected to give a time estimate within the same order of magnitude as if the simulations were done with MD.
\begin{figure*}
 \centering
\subfigure[Initial]{
  \includegraphics[width=0.3\textwidth]{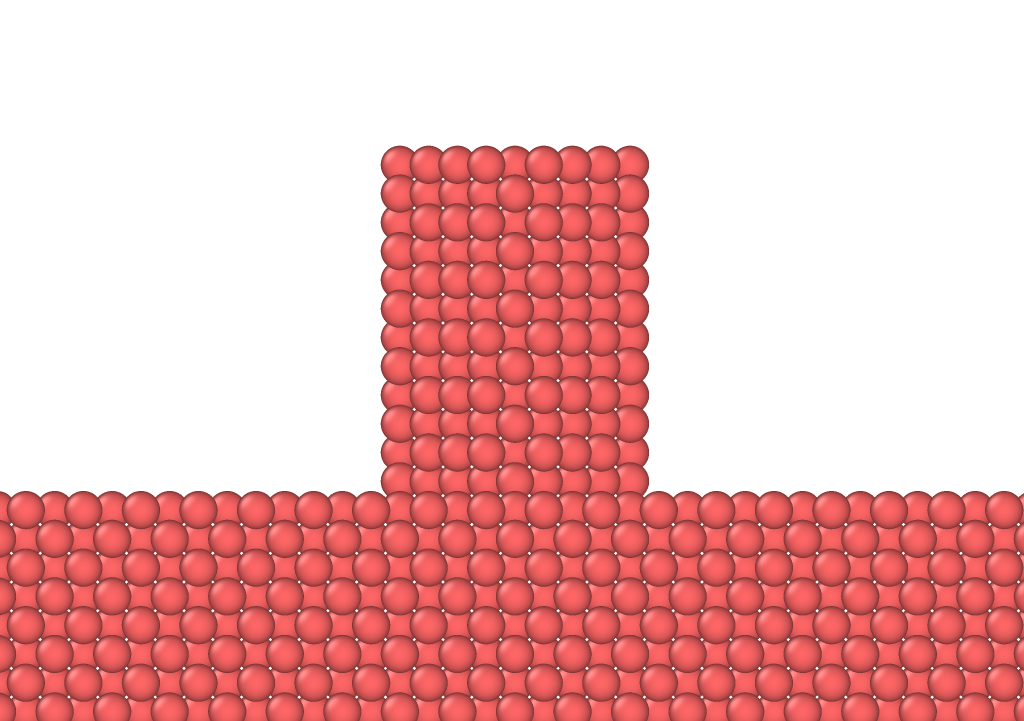}\label{fig:cylinder_a}
}
\subfigure[MD]{
  \includegraphics[width=0.3\textwidth]{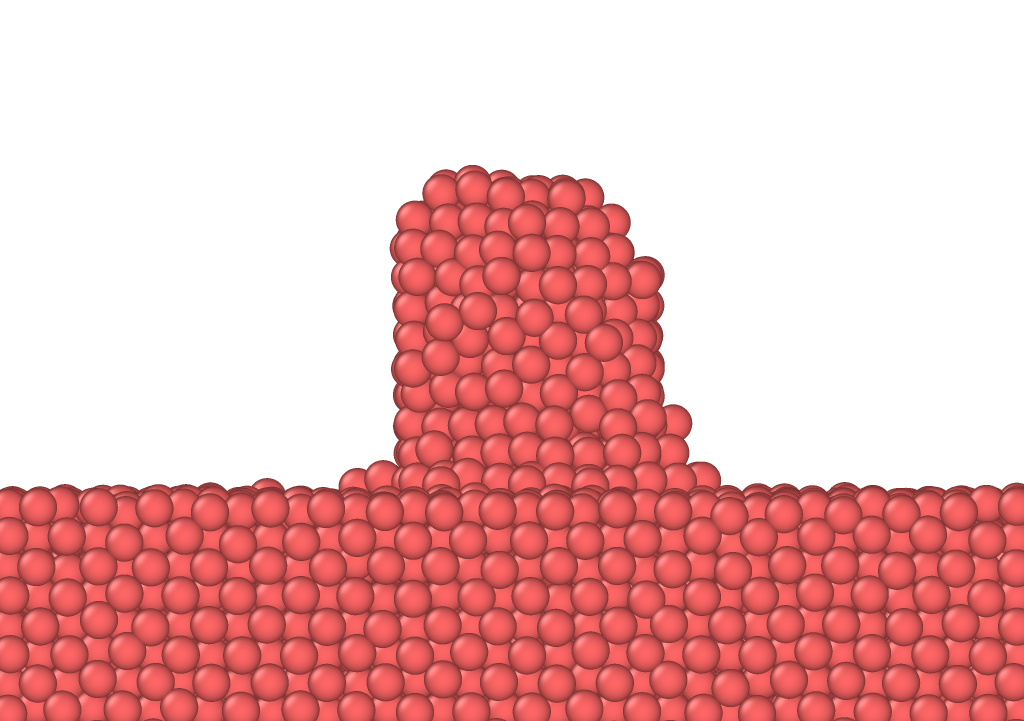}
}
\subfigure[KMC]{
  \includegraphics[width=0.3\textwidth]{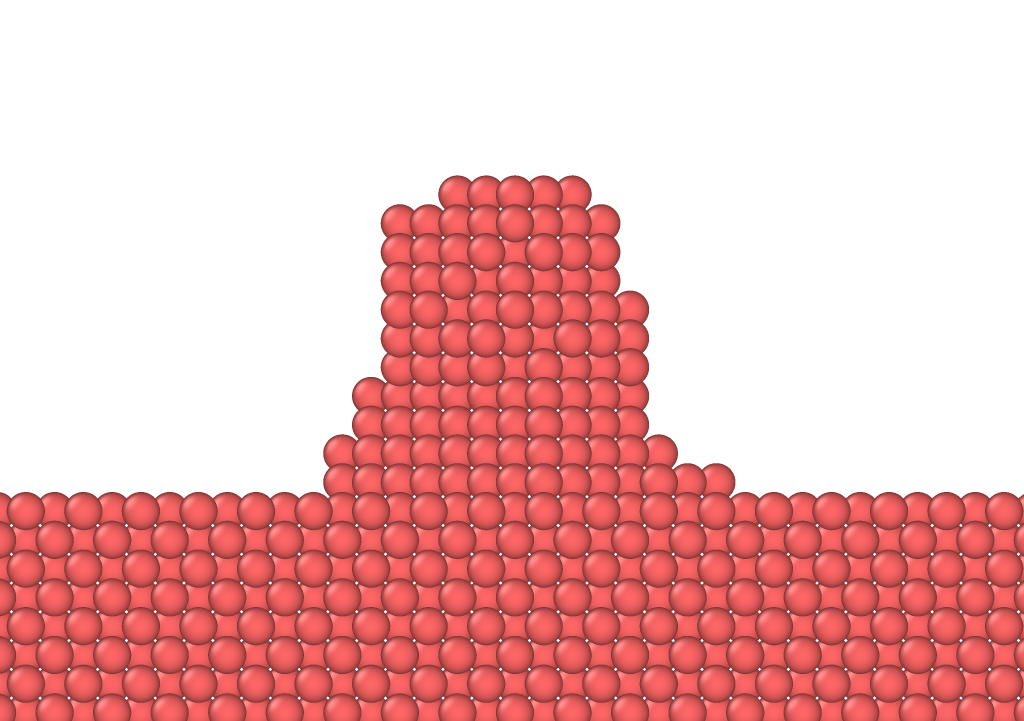}
}

\caption{Flattening of a W cylindrical nanotip (a), shown as a cross-section. The final shape is shown for MD in (b), and for our KMC model in (c). The figures are made with Ovito \cite{ovito}.}
\label{fig:cylinder}
\end{figure*}

The MD simulations were done at constant volume and temperature, 3000\;K.
This is still only 2/3 of the melting temperature of the interatomic potential (EAM4 \cite{marinica2013interatomic}), which is 4600\;K \cite{sand2016non}, and should be low enough to avoid any non-linear changes of the migration barriers at higher temperatures \cite{bukonte2014modelling,mundy1987vacancy}.
The W nanotip was a 2.7\;nm high cylinder with a 1.8\;nm diameter, standing on a bcc\{110\} substrate, $10\times10$\;nm$^{2}$, and 2.9\;nm thick.
The lowest layer of atoms were fixed. 
The lattice parameter of the atomic system was obtained in a separate simulation by relaxing the volume of a bulk system at constant zero pressure at the same 3000\;K temperature.

Tungsten is a fairly rigid material and the W tip will therefore be rather stable, but at 3000\;K the tip will still slowly decrease in height due to atom diffusion, while maintaining its crystal lattice structure.
A similar tip flattening has also been experimentally observed in W nanowires when a high field emission current density runs through them \cite{yeong2006field}.
The flattening time is here defined as the time it takes for the tip to decrease in height below a threshold that corresponds to a decrease of one atomic layer (figure \ref{fig:md_height}).
The average flattening time was, after 10 repeated runs with different seeds, found to be $(3.02 \pm 0.98) \cdot 10^{-9}$\;s.
\begin{figure}
 \centering
 \includegraphics[width=\columnwidth]{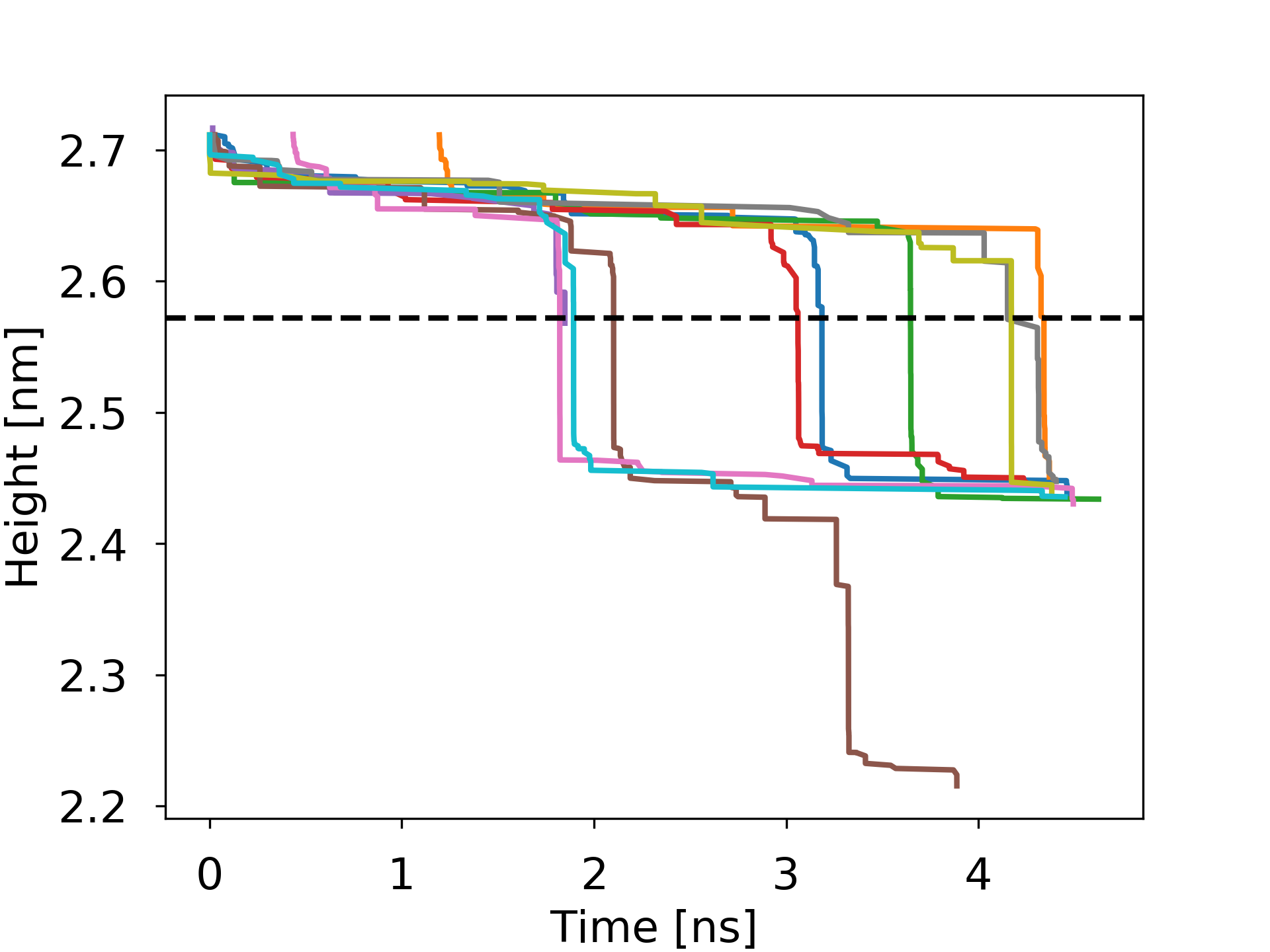}
 \caption{MD simulations of the W nanotip flattening, showing the nanotip height versus simulation time for ten different statistical runs.
 The height evolution is seen to be discrete as the bcc lattice of the nanotip is largely intact.
 When all atoms are below the height threshold 2.57\;nm, marked with a dashed line, the top atomic monolayer of the nanotip is empty.}
 \label{fig:md_height}
\end{figure}

The same system was subsequently simulated with KMC using ten different attempt frequencies, ten different runs for every case, and the best fit attempt frequency $\nu = (4.3 \pm 2.1) \cdot 10^{14}$\;s$^{-1}$ could be found by comparing the flattening times to the MD value (figure \ref{fig:nu_fit}), which we can now use as our global attempt frequency in our KMC model.
The nanotips in the KMC simulations followed the same qualitative evolution as observed in the MD simulations and can be seen in figure \ref{fig:cylinder}.
\begin{figure}
 \centering
 \includegraphics[width=\columnwidth]{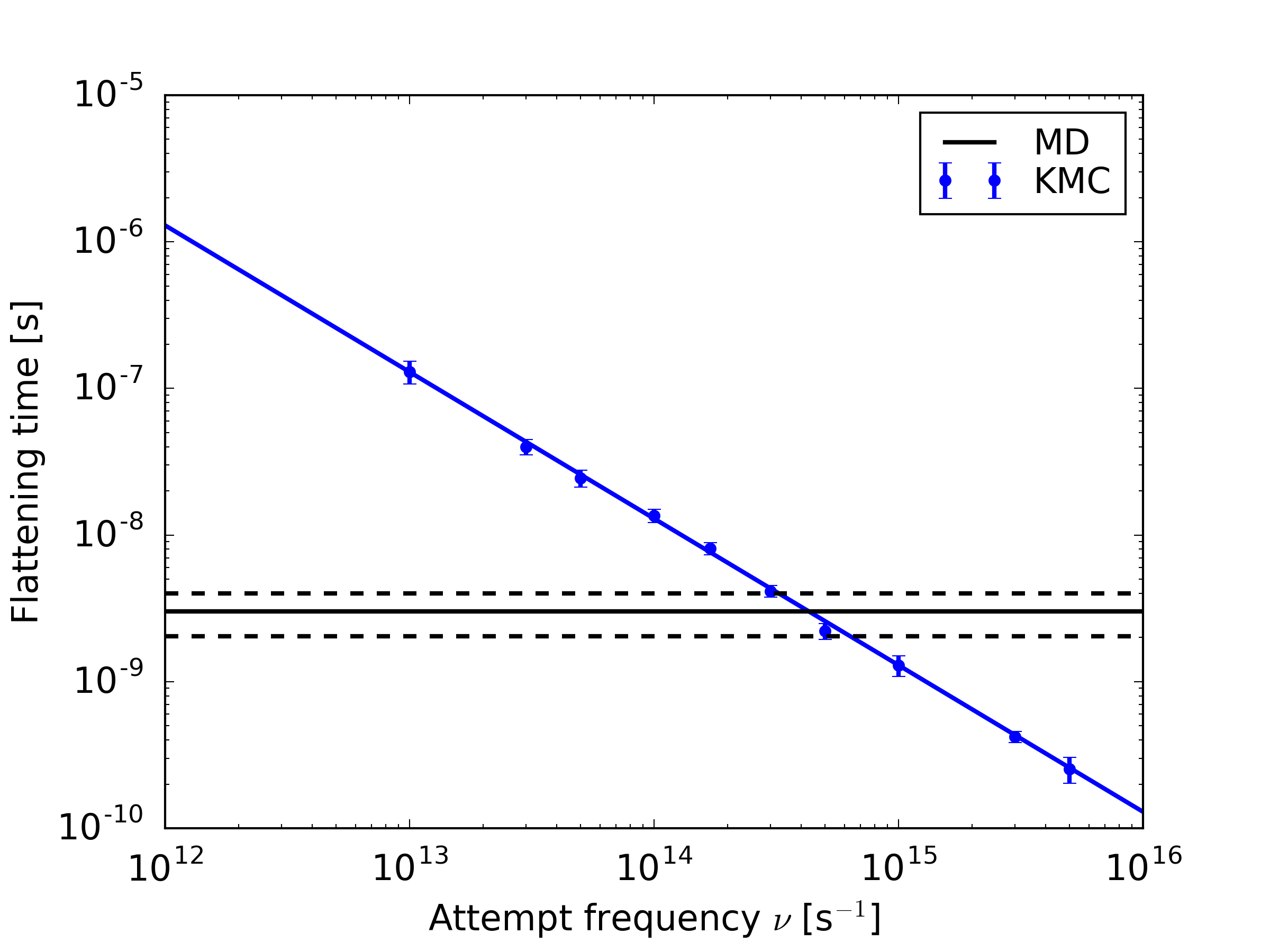}
 \caption{KMC simulations of the flattening of the W nanotip with different attempt frequencies $\nu$.
 The MD value for the flattening time and its errobars are shown with black full and dotted lines, respectively.
 The intersection between the KMC and the MD results is found at $\nu = (4.3 \pm 2.1) \cdot 10^{14}$\;s$^{-1}$.
 The errorbars are the standard deviations of ten different statistical runs.
 }
 \label{fig:nu_fit}
\end{figure}

\subsection{Validation of the W parameterization}\label{sec:wulff}

% R20190507_wulff_dft  
% R20190507_wulff_marinica

We validated the W parameterization by simulating four different atom clusters, labelled I to IV and shown in figure \ref{fig:wulff}.
Any W cluster of any shape should with enough time evolve towards its energy minimum shape, which theoretically should be quite close to the Wulff construction shape.
The Wulff construction is based on the surface energies of the different possible facets.  
In our case, as the barriers are calculated using an interatomic potential, the minimum energy shape should be the Wulff construction based on the surface energies calculated with the same interatomic potential.
These values are listed in table \ref{table:surface_energies}.
The Wulff construction will also vary in shape with the size of the cluster.
For creating our Wulff-constructed clusters, we used the Atomic Simulation Environment Python library \cite{ase}.

We simulated with KMC four different W clusters (I--IV) to see if their thermal evolution would approach the Wulff construction for the respective sizes.
Cluster I and II had both 1024 atoms and a temperature of 3000\;K.
For comparison, we also simulated with MD cluster I and II at the same temperature in order to compare the time evolution.
Cluster I was initially in the shape of the Wulff construction [figure \ref{fig:Ia}]. 
After simulating with KMC for 17\;ns or $10^6$ steps (or atom transitions), the cluster was still largely unchanged [figure \ref{fig:Ib}], in good agreement with the MD simulation of the same cluster for the same time [figure \ref{fig:Ic}].
The KMC simulations was continued until $2.0\cdot10^{-7}$\;s or $1.2\cdot10^7$ steps had passed with no signifcant change, as expected. 
Cluster II had the initial shape of a cube [figure \ref{fig:IIa}]. 
Both the KMC [figure \ref{fig:IIb}] and the MD [figure \ref{fig:IIc}] simulations of the cluster became very close to the Wulff construction already after $1.7\cdot10^{-8}$\;s ($10^6$ KMC steps). The KMC simulation was continued until $2.0\cdot10^{-7}$\;s or $1.2\cdot10^7$ steps had passed with no significant change. 

Clusters III and IV had both 39\,000 atoms and a diameter of $2\cdot10^{-7}$\;m and a temperature of 2300\;K.
Both cluster were given the shapes of the Wulff construction, but cluster III used the surface energies calculated with the interatomic potential and cluster IV the surface energies calculated with DFT.
All surfaces energies are listed in table \ref{table:surface_energies}.
The MD values are taken from \cite{marinica2013interatomic} and the DFT values from \cite{vitos1998surface}.	
Both the MD potential and DFT favours the \{110\} surface, but the \{100\} surface is slightly more favoured, compared to the other surfaces, by the MD potential than by the DFT calculations, as can be seen in table \ref{table:surface_energies}.
\begin{figure*}
 \centering
\subfigure[I, initial]{
  \includegraphics[width=0.3\textwidth]{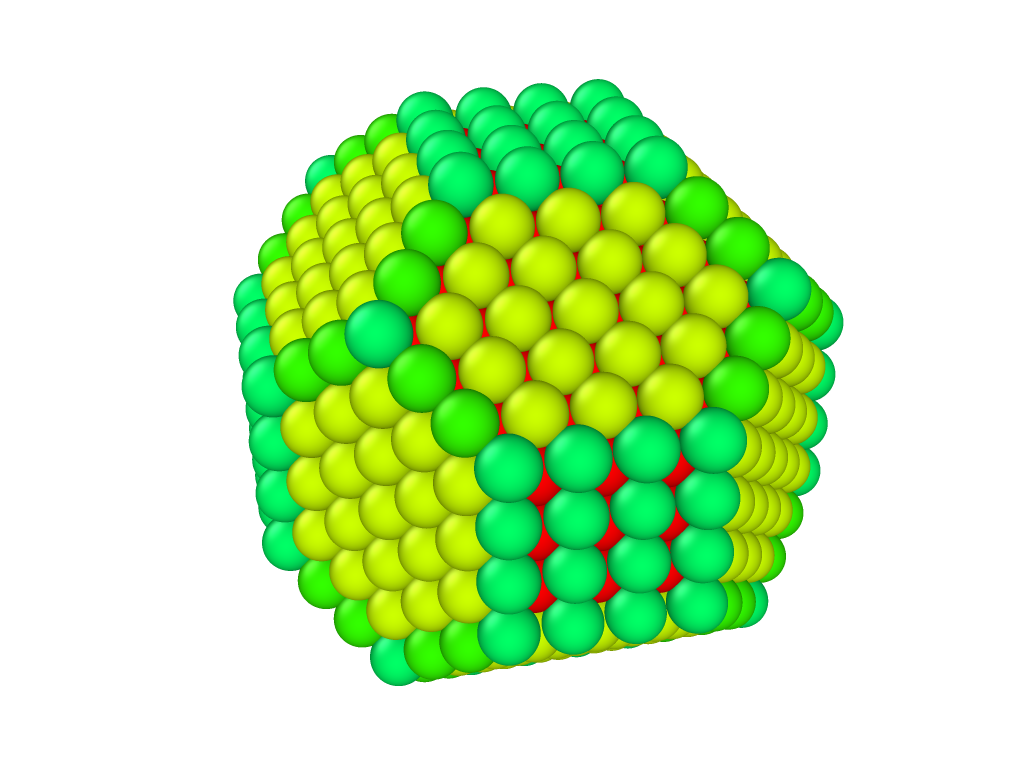}\label{fig:Ia}
}
\subfigure[I, KMC, $t = 17$\;ns]{
  \includegraphics[width=0.3\textwidth]{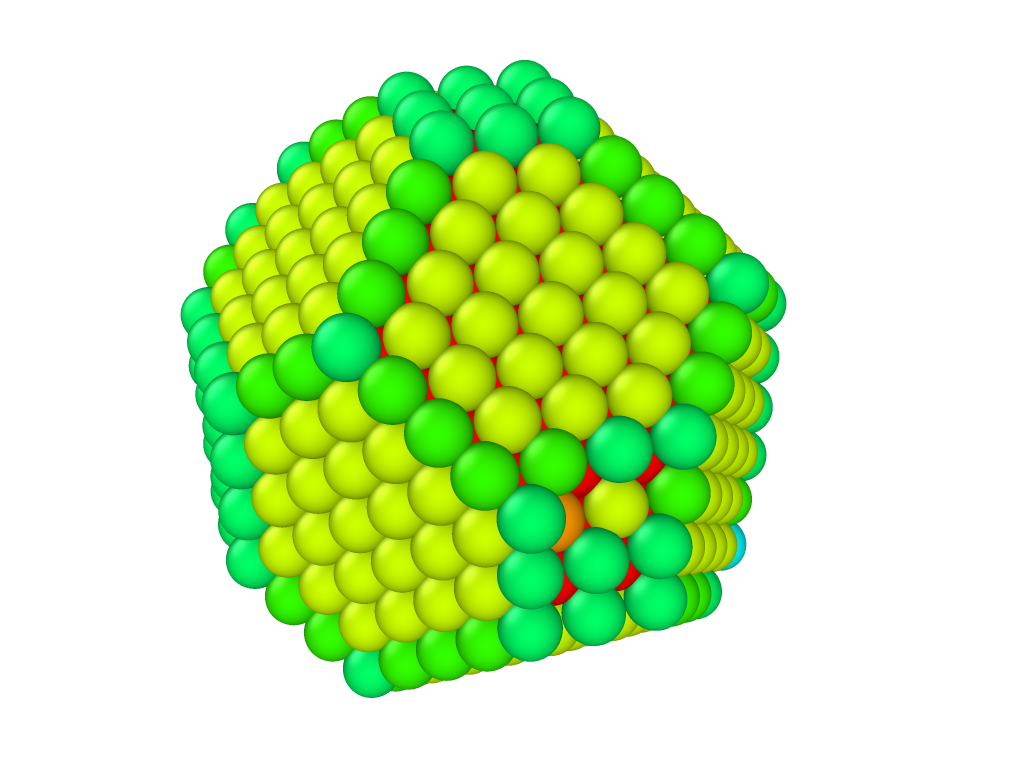}\label{fig:Ib}
}
\subfigure[I, MD, $t = 17$\;ns]{
  \includegraphics[width=0.3\textwidth]{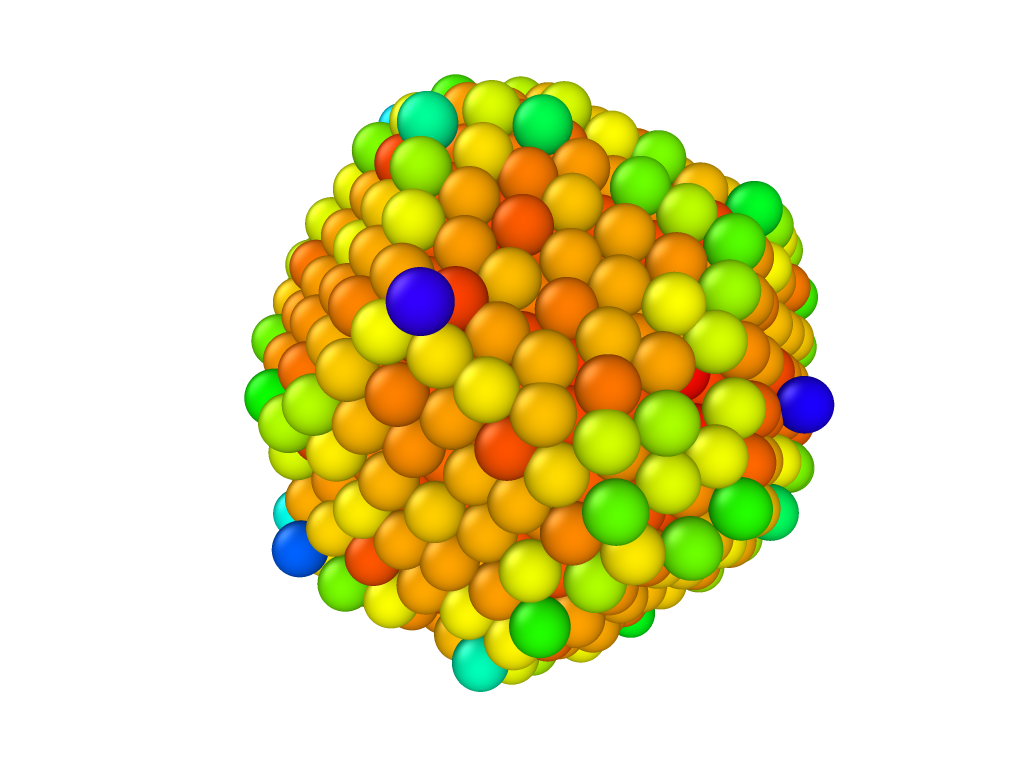}\label{fig:Ic}
}\\
\subfigure[II, initial]{
  \includegraphics[width=0.3\textwidth]{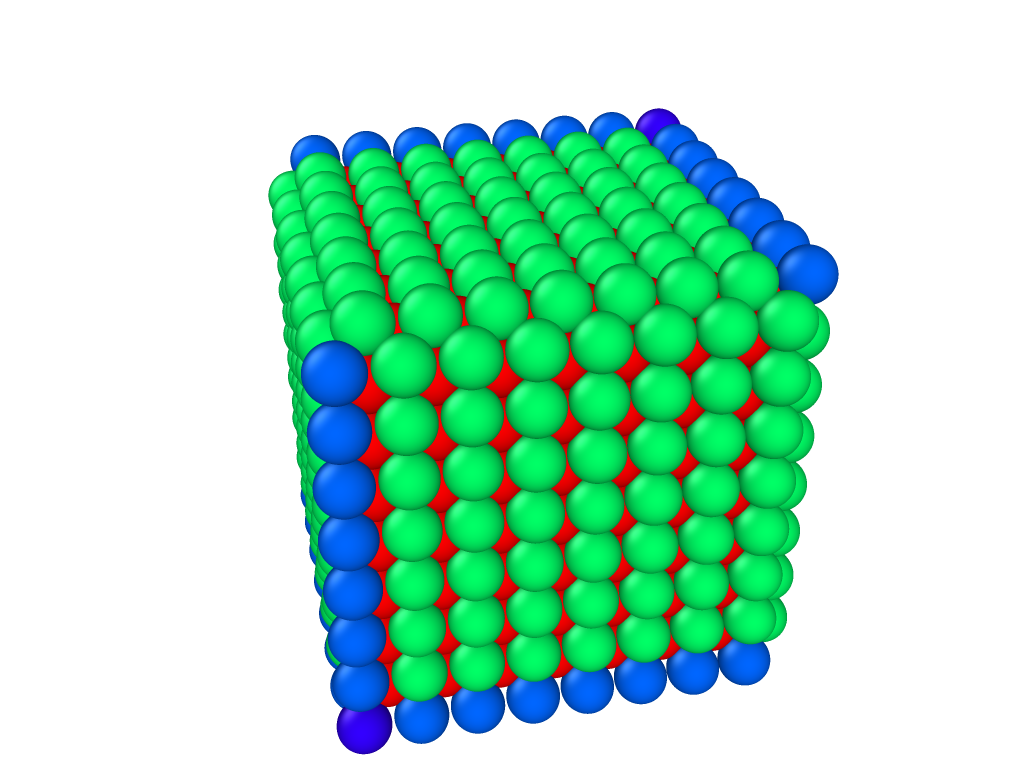}\label{fig:IIa}
}
\subfigure[II, KMC, $t = 17$\;ns]{
  \includegraphics[width=0.3\textwidth]{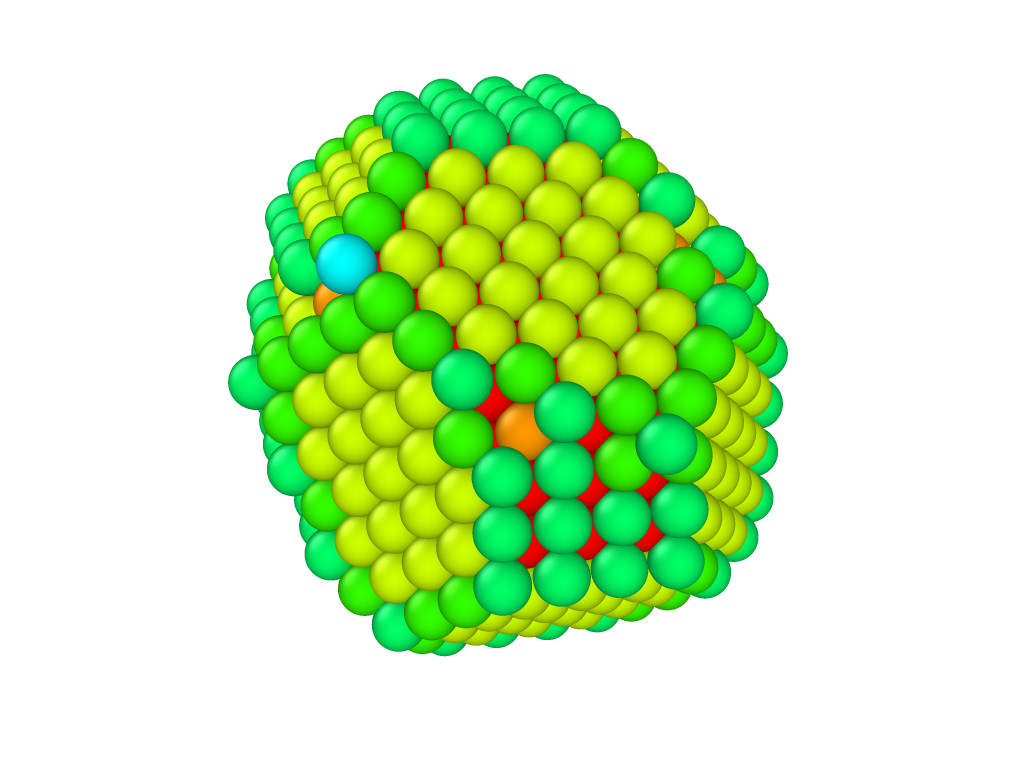}\label{fig:IIb}
}
\subfigure[II, MD, $t = 17$\;ns]{
  \includegraphics[width=0.3\textwidth]{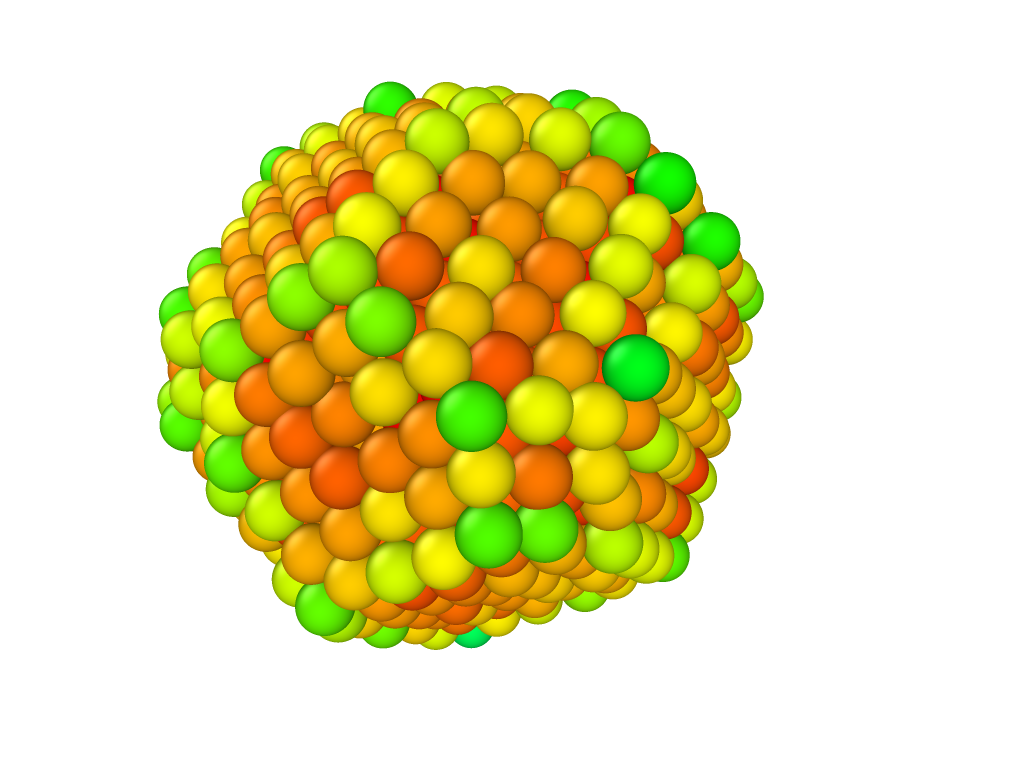}\label{fig:IIc}
}\\
\subfigure[III initial]{
  \includegraphics[width=0.7\columnwidth]{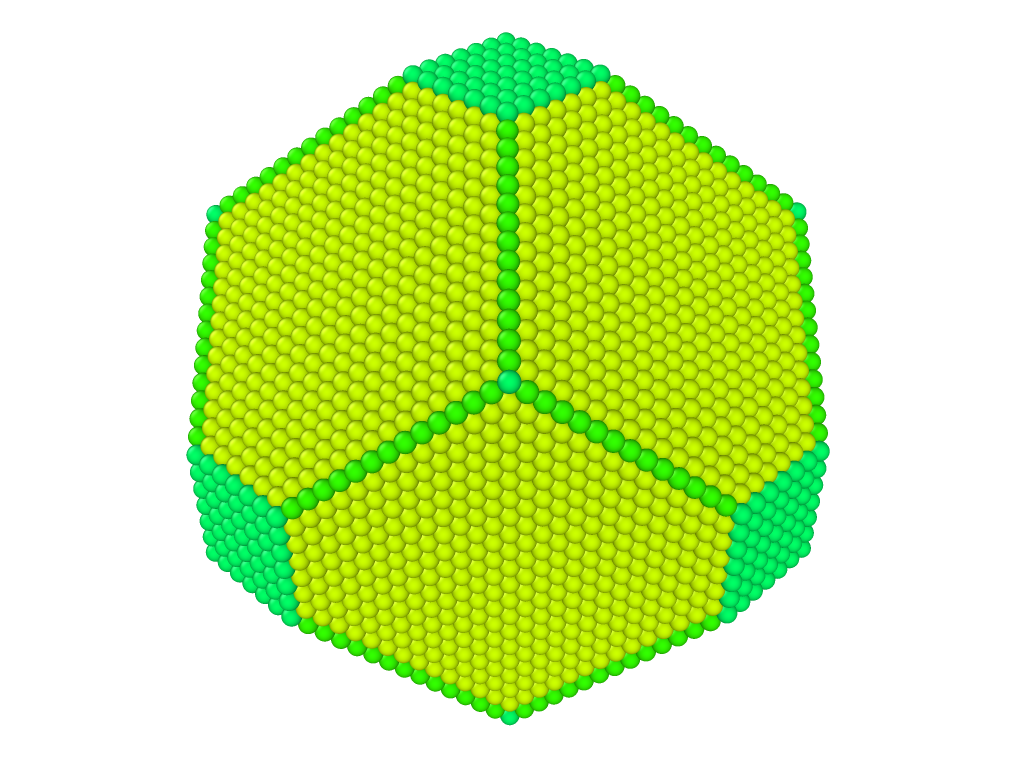}\label{fig:wulff_marinica_initial}
  % width = 0.4
}
\subfigure[III, KMC, $t=6.2$\;$\mu$s]{ %6.184687e-06 s
  \includegraphics[width=0.7\columnwidth]{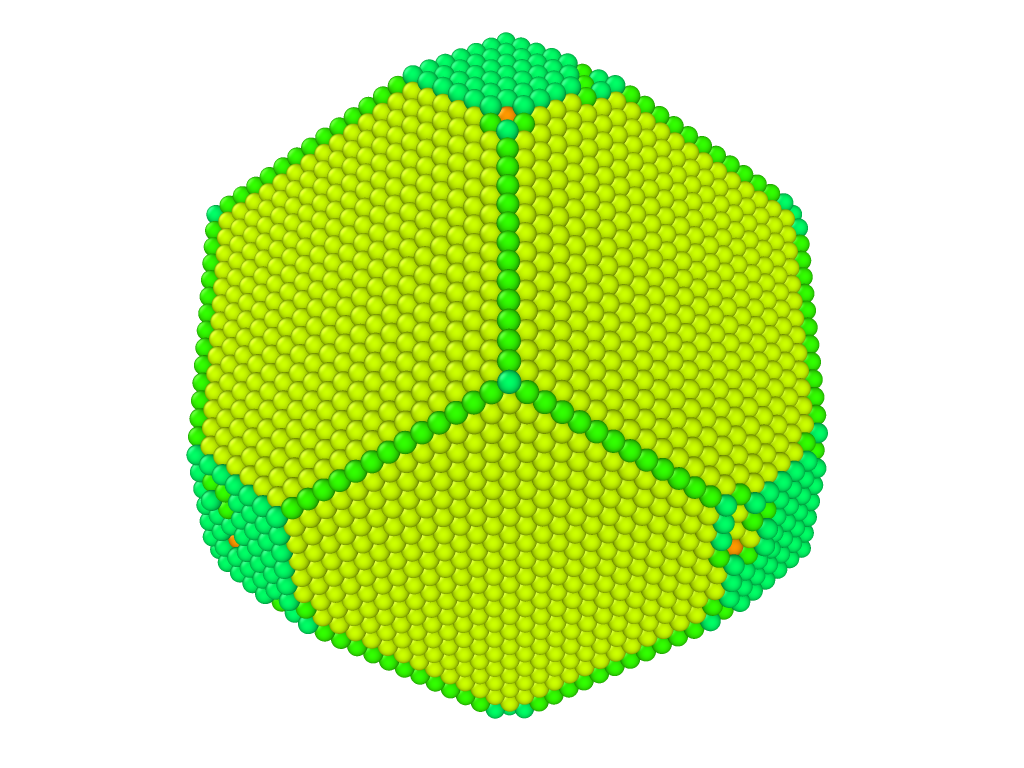}\label{fig:wulff_marinica_end}
}\\
\subfigure[IV initial]{
  \includegraphics[width=0.7\columnwidth]{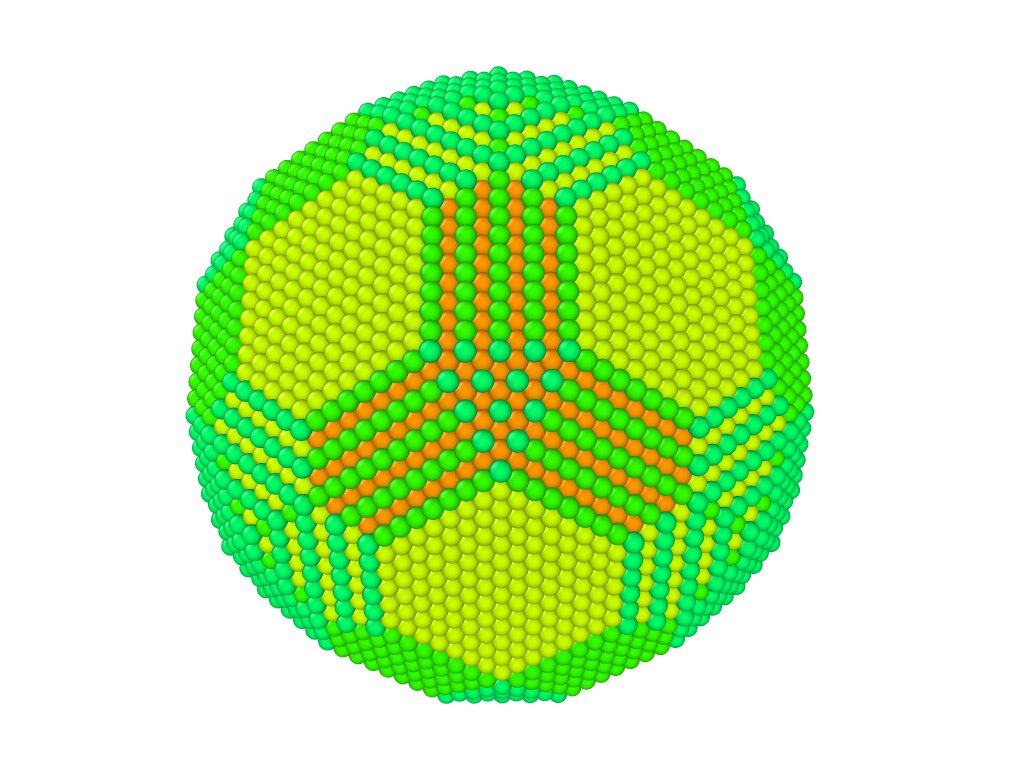}\label{fig:wulff_dft_initial}
}
\subfigure[IV, KMC, $t=6.5$\;$\mu$s]{ %6.548817e-06 s 
  \includegraphics[width=0.7\columnwidth]{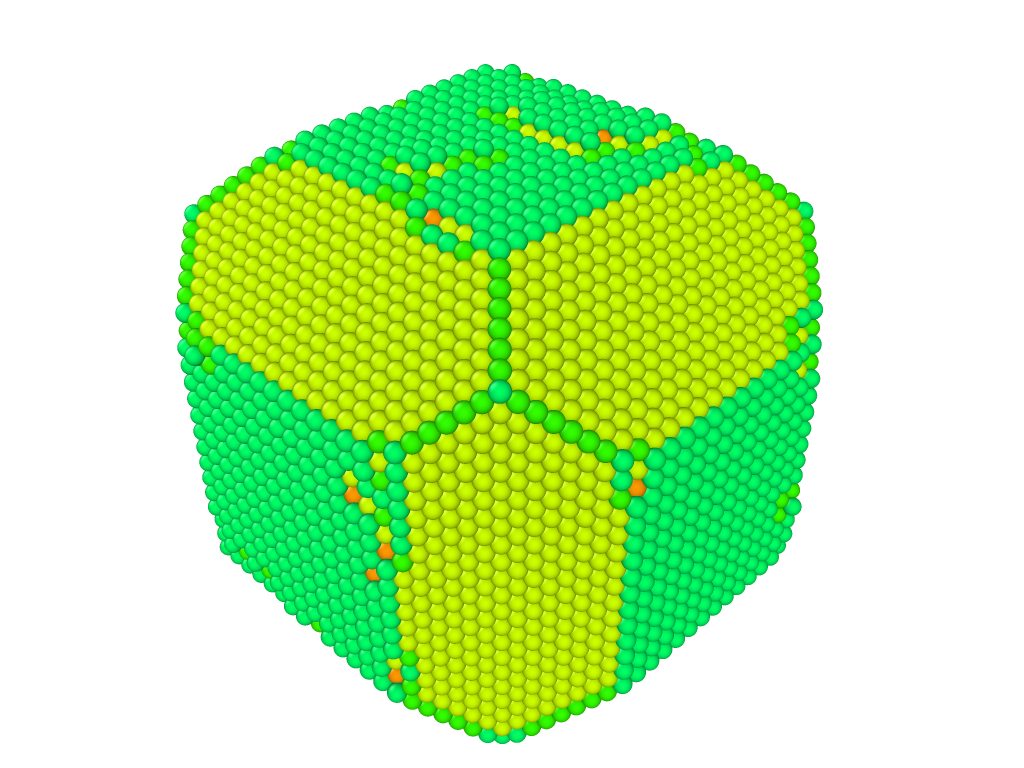}\label{fig:wulff_dft_end}
}
\caption{KMC and MD simulations of W clusters, showing the initial and final shapes. Cluster I and II have 1024 atoms and are simulated at 3000\;K. Figures (a--c) show the intial shape, the final KMC shape, and the final MD shape, respectively. Cluster I has initially (a) the shape of the Wulff construction using the MD potential surface energy values. (d--f) show likewise the inital shape, the final KMC shape, and the final MD shape for Cluster II, respectively. Cluster III and IV have 39\,000 atoms and are simulated at 2300\;K. Both cluster III (g) and IV (i) have the initial shapes of the Wulff construction, but III is using the MD potential surface energy values and IV the DFT values. (h) and (j) show the final shapes of cluster III and IV, respectively, after 6\;µs. The atoms in the KMC simulations are coloured according to their coordination number in order to highlight the different facets. \{110\} facets are shown as yellow, \{100\} green, \{211\} facets are red-green striped, and \{310\} facets are yellow-green striped. 
For the MD simulations, the facets are highlighted by colouring the atoms according to their centrosymmetry.
The figures are made with Ovito \cite{ovito}.}
\label{fig:wulff}
\end{figure*}
\begin{table}
  \centering
   \caption{Surface energies for different W bcc surfaces, relative to the energy of the \{110\} surface energy $\gamma_{110}$, which with MD is 15.7\;eV\;nm$^{-2}$ (2.52\;J\;m$^{-2}$) and with DFT 25.0\;eV\;nm$^{-2}$ (4.005\;J\;m$^{-2}$). The MD values are from \cite{marinica2013interatomic} and the DFT values from \cite{vitos1998surface}.}
   \label{table:surface_energies}
   \begin{tabular*}{\columnwidth}{@{\extracolsep{\fill}} l c c}
   \toprule
              & MD   & DFT \\
   $\{ijk\}$ & $\gamma_{ijk}/\gamma_{110}$ & $\gamma_{ijk}/\gamma_{110}$\\
   \midrule
   \{110\} &  1.000 & 1.0000 \\
   \{100\} &  1.165 & 1.1573 \\
   \{211\} &  1.189 & 1.0429 \\
   \{310\} &  1.191 & 1.0744 \\
   \{311\} &  1.231 &        \\
   \{511\} &  1.233 &        \\
   \{111\} &  1.278 & 1.1116 \\
   \bottomrule
\end{tabular*}
\end{table}

The most stable cluster was, as expected, cluster III [figures \ref{fig:wulff_marinica_initial} and \ref{fig:wulff_marinica_end}] with the initial shape of the Wulff construction based on MD values (table \ref{table:surface_energies}), as the KMC barrier set were also based on this potential. 
The Wulff shape features large \{110\} (yellow) and small \{100\} facets (green).
In the case of cluster III, the shape was still unchanged after 6.2\;µs or $1.1\cdot10^8$ steps.
In KMC, only one transition (or event) happens every step.
For most of the $10^8$ steps, a first-nearest neighbour jumps was carried out, but $9.6\cdot10^4$ second-nearest jumps and $4.5\cdot10^5$ exchange processes were also carried out during the simulation.

Cluster IV [figures \ref{fig:wulff_dft_initial} and \ref{fig:wulff_dft_end}] had as the initial shape the Wulff construction based on the DFT surface energy values (table \ref{table:surface_energies}).
The Wulff shape has larger \{211\} (red-green) and \{310\} facets (green-yellow), and smaller \{110\} (yellow) facets than cluster III.
After 2.5\;µs, the \{211\} and \{310\} facets have disappeared and large \{100\} facets have formed and this shape stays until the end at 6.5\;µs or $1.1\cdot10^8$ steps, as shown in figure \ref{fig:wulff_dft_end}, suggesting that the large \{100\} facets form a local energy minimum, even though it can not be excluded that cluster IV, given enough time, would transform to the same shape as cluster III, which should have a shape very close to the global minimum given by the MD potential that the KMC parameter set is based on.
Of all $10^8$ transitions in the simulation, $4.0\cdot10^5$ were second-nearest neighbour jumps and $1.5\cdot10^6$ third-nearest neighbour exchange processes.

\section{Discussion}\label{sec:discussion}

In this work, we have calculated a parameterization for W surfaces in a similar manner as were done previously for Cu, Fe and Au in \cite{baibuz2018migration,vigonski2018au}, with a near-complete set (pure bulk processes not considered) of second-nearest neighbour jumps included and also a third-nearest neighbour exchange process.
The parameterization describes a set of all atom transitions that we consider necessary for KMC simulations of W surfaces.
For every process, labelled by its $(a,b,c,d)$ configuration (see section \ref{sec:methods}), a migration barrier has been calculated with NEB.
We have calculated three separate subsets, which together constitute the self-sufficient W parameterization:
\begin{itemize}
  \item A complete subset of first-nearest neighbour atom jumps.
  \item A near complete subset of second-nearest neighbour atom jumps. Processes with almost all neighbours present were omitted to limit the number of NEB calculations (see section \ref{sec:parameterization}). These processes are anyway unlikely to happen in near-bulk conditions, as the neighbours will likely block the transition path.
  \item A subset with the single third-nearest neighbour distance (4,1,4,1) exchange process; included as it plays a significant role on the bcc\{100\} surface. 
  Exchange processes are more complicated than jump processes as at least two atoms are involved, which also increases the number of possibilities for the reaction paths of the atoms for a particular process. 
  To construct a complete set of exchange processes is therefore beyond the scope of this work and we have settled for only including the most simple third-nearest neighbour distance exchange (4,1,4,1) process, which, because of its relatively low barrier, 1.95\;eV, is likely to be important.
\end{itemize}

Of the 1734 first-nearest neighbour jump processes, 100 barriers were calculated explicitly on the closed-packed \{110\} surface and 26 on the other important \{100\} surface.
Because of the complicated structure of the \{111\} surface, only 5 barriers could be calculated on this surface.
Of all $(a,b,c,d,)$ jump labels, there were no overlap between the \{110\} and \{100\} surface, which means that the model should be able to distinguish between these two surfaces accurately.
For the \{111\} surface, all but one process label was also found in the \{110\} set, but since these are very few barriers and the \{111\} surface anyway is one of the least energetically favourable surfaces, it is a reasonable approximation to ignore this small overlap. 
In the final barrier set, the barrier values from the \{110\} and \{100\} surfaces have been chosen if a duplicate barrier label have been found in the \{111\} surface or in the bulk calculations, were all $(a,b,c,d)$ possibilities were gone trough systematically (as described in section \ref{sec:neb}).

The barriers are mostly below 6\;eV, as can be seen in figure \ref{fig:barriers}, but a few barriers can be seen to be around 7 and 8\;eV.
These outlier barriers may or may not be artefacts from the automatic calculations of the barriers.
In any case, a few barriers with very high barriers are unlikely to change the overall behaviour of the whole barrier set of thousands of barriers.

In table \ref{table:barriers}, the energy barriers for three processes (first-, second-nearest neighbour jump, and a exchange process) are shown.
It can be seen that the effect of the tethering force for these processes are 0.02\;eV at most, which is minimal.
In \cite{baibuz2018migration}, it was found for the generally smaller Cu barriers, that 95 \% of the barriers were changed by less than 0.5 eV by the tethering force, which is still small.
Comparing the non-tethered NEB calculated barriers in table \ref{table:barriers} with DFT values from the literature, listed in the same table, suggests that the interatomic potential by Marinica et al. \cite{marinica2013interatomic} may predict the barriers with a precision of 0.2--0.4\;eV, although the order of the barrier values are the same as calculated with DFT.
It can also be noted that the Marinica potential does not predict the ``zigzag'' reconstruction of the W\{100\} at temperatures below 250\;K, which has been observed in experiments and DFT calculations \cite{debe1977space,heinola2010first,olewicz2014coexistence}, and would give slightly higher barrier values, as shown in table \ref{table:barriers}. 
The potential also slightly underestimate the surface energy of the \{100\} surface, compared to DFT, as shown in table \ref{table:surface_energies}.
The other surface energies, as calculated by the potential, follows the same order as found in DFT, which is more essential for capturing the correct surface dynamics than having precise values for the surface energies.
The potential by Derlet et al. \cite{derlet2007multiscale} have all the surface energies in the same order as DFT, but the NEB calculations with this potential were found to converge too slowly for our needs, as thousands of NEB calculations are needed for a complete barrier set. 
With these considerations, the potential by Marinica \cite{marinica2013interatomic} was considered the best choice.
Systematic errors in the barrier values may not necessarily affect the dynamics of the KMC simulations, but might affect the time estimates.
This may to some degree be compensated by the fitting of the attempt frequency (section \ref{sec:attempt}), which ensures that the time scale is comparable to corresponding MD simulations.

Our estimations, using two different methods, of a second-nearest neighbour atom jump in section \ref{sec:attempt} did not find support for G\;Antczak and G\;Ehrlich, claims that such processes would have four orders of magnitudes higher attempt frequencies. 
On the contrary, we found a difference between first- and second-nearest neighbour jumps of at most one order of magnitude, using two different methods: MD and harmonic approximation. 
The attempt frequency of the single third-nearest neighbour exchange process included in the parameterization was estimated with MD to be within the same order of magnitude, $10^{12}$ s$^{-1}$, as the first-nearest neighbour jump.
The exchange process was not estimated with harmonic approximation as this method may not be valid for other than single jump transitions.
Since we can expect the attempt frequencies of all transitions to be relatively close to each other and as calculating all thousands of attempt frequencies would be quite challenging, it is well motivated to fit and use a global attempt frequency $\nu$ for all atom transitions in the parameterization.
The global attempt frequency $\nu = (4.3 \pm 2.1) \cdot 10^{14}$\;s$^{-1}$ was fitted by simulating the thermal flattening of small 2.7\;nm high W tips at 3000\;K and comparing with the corresponding MD simulations, as described in section \ref{sec:attempt}.
Similar fitting of global attempt frequencies has been done in previous KMC works \cite{jansson2016long, baibuz2018migration, vigonski2018au}.
This $\nu$ will only act as a scaling factor for the time estimates in the KMC model and only roughly correspond to the attempt frequencies that were earlier estimated for individual transitions in table \ref{table:barriers}.
The relatively large difference in value between the calculated attempt frequencies and the fitted $\nu$ may come from some barrier values that are calculated imprecisely in the automated NEB calculations. 
Since we have thousands of barriers, it is hard to pinpoint individual problematic barriers.
It is also possible that adding more long transitions, such as third-nearest exchange processes, may improve the precision of the model, but these processes may require some more thorough systematic studies, which we will leave for the future. 
Instead we can look at the barrier set as a whole and see that it produces the correct dynamics, as was done in section \ref{sec:wulff}.
Since the same $\nu$ is used for all transitions, the dynamics of the simulations will not be affected by the value of $\nu$.
KMC simulations of W systems have timescales that may be vastly larger than MD simulations, which makes it difficult to make corresponding MD simulations for comparing the time scales. 
Our value of $\nu$ is fitted to flattening process of W nanotips and can therefore be expected to give roughly correct time scales for KMC simulations of other W nanotips.
How the time estimate given by our value of $\nu$ would be affected by other surface processes than nanotip flattening, is beyond the scope of this work.

In section \ref{sec:wulff}, we showed that W clusters simulated with KMC and our parameterization will evolve towards the energetically favourable shape predicted by the Wulff construction.
The MD simulations of the same clusters at the same temperature showed the same evolution, which is a strong validation of the parameterization.
Since these kind of simulations can be be very time consuming and it is hard to tell at which point a particular cluster has found an equilibrium shape and stopped evolving, we also used an approach where we started with the Wulff construction and showed that the cluster will not change over a reasonable large number of KMC steps. 
For the small clusters with 1024 atoms we could show that a cubic clusters evolved as predicted into the Wulff construction shape. 
If the cluster started in the Wulff construction shape, it would not evolve at all for the same number of steps.
The MD simulations showed the same evolution.
A larger W cluster with 39\;000 atoms was simulated at a lower temperature, which started in the shape of the Wulff construction and did indeed not significantly evolve after 6.2\;µs or $10^8$ steps, which corresponds to four weeks of CPU time. 
We also tried a slightly different initial shape, actually based on the Wulff construction using the DFT surface energy values, with the expectation that this cluster would evolve into the same shape as the other large cluster, but it seems that $10^8$ steps were not enough for this to happen and it is possible that the cluster found a local energy minimum shape, which given enough time, still would evolve towards the Wulff construction shape, predicted by the potential.
The fact that all clusters with the shape of the Wulff construction were found stable and that the small cubic cluster evolved into the Wulff shape, in good agreement with the MD simulations, are good confirmations of the validity of the KMC model and our W parameterization.

\section{Conclusions}\label{sec:conclusions}

We have calculated the surface migration barriers that form a self-sufficient parameterization suitable for Kinetic Monte Carlo simulations of arbitrary rough tungsten (W) surfaces and nanostructures, such as e.g. nanotips, or nanoclusters. 
The parameterization includes first- and second-nearest neighbour atom jump barriers.
A third-nearest neighbour exchange process barrier for the \{100\} surface was also calculated.
An attempt frequency was fitted to give a time scale of the same order of magnitude as in Molecular Dynamics simulations.
The same attempt frequency was found sufficiently precise for any first-, second- or third-nearest neighbour atom transition. 
The parameterization was validated by correctly simulating the Wulff-constructed energy minimum cluster shapes in good agreement with Molecular Dynamics simulations.

\ack
V\;Jansson was supported by Academy of Finland (Grant No.\;285382) and Waldemar von Frenckells Stiftelse.
A\;Kyritsakis were supported by a CERN K-contract (No.\;47207461).
E\;Baibuz was supported by the CERN K-contract (No.\;47207461) and the doctoral school DONASCI of the University of Helsinki.
The work of V\;Zadin and A\;Aabloo was supported by Estonian Research Council Grants PUT1372 and IUT20-24.
F\;Djurabekova acknowledges gratefully the financial support of Academy of Finland (Grant No.\;269696).
The authors would like to thank Dr Tommy Ahlgren and Dr Antti Kuronen for fruitful discussions.
Computing resources were provided by the Finnish IT Center for Science (CSC) (persistent identifier urn:nbn:fi:research-infras-2016072533).

\section*{ORCID iDs}
{\small
V\;Jansson     \url{https://orcid.org/0000-0001-6560-9982}\\
A\;Kyritsakis  \url{https://orcid.org/0000-0002-4334-5450}\\
S\;Vigonski    \url{https://orcid.org/0000-0002-2849-2882}\\
E\;Baibuz      \url{https://orcid.org/0000-0002-9099-1455}\\
V\;Zadin       \url{https://orcid.org/0000-0003-0590-2583}\\
A\;Aabloo      \url{https://orcid.org/0000-0002-0183-1282}\\
F\;Djurabekova \url{https://orcid.org/0000-0002-5828-200X}
}

\bibliographystyle{my-iopart-num.bst}
\bibliography{pub/vjansson.bib,pub/vjansson_publications.bib}

\providecommand{\newblock}{}
\providecommand{\url}[1]{{\tt #1}}
\providecommand{\urlprefix}{}
\providecommand{\href}[2]{#2}
\begin{thebibliography}{10}

\bibitem{boling1958blunting}
Boling J~L and Dolan W~W 1958 \href{https://doi.org/10.1063/1.1723220}{{\em
  Journal of Applied Physics\/} {\bf 29} 556--9}

\bibitem{barbour1960determination}
Barbour J, Charbonnier F, Dolan W, Dyke W, Martin E and Trolan J 1960 {\em
  Physical Review\/} {\bf 117} 1452

\bibitem{tsong1975direct}
Tsong T and Kellogg G 1975 {\em Physical Review B\/} {\bf 12} 1343

\bibitem{wang1982field}
Wang S and Tsong T 1982 {\em Physical Review B\/} {\bf 26} 6470

\bibitem{yeong2006field}
Yeong K~S and Thong J~T~L 2006 \href{https://doi.org/10.1063/1.2400722}{{\em
  Journal of Applied Physics\/} {\bf 100} 114325}

\bibitem{fujita2007mechanism}
Fujita S and Shimoyama H 2007 {\em Physical Review B\/} {\bf 75} 235431

\bibitem{suzuki2018single}
Suzuki Y and Kizuka T 2018 {\em Applied Physics Express\/} {\bf 11} 045201

\bibitem{vurpillot2018simulation}
Vurpillot F, Parviainen S, Djurabekova F, Zanuttini D and Gervais B 2018 {\em
  Materials Characterization\/} {\bf 146} 336--46

\bibitem{vigonski2018au}
Vigonski S, Jansson V, Vlassov S, Polyakov B, Baibuz E, Oras S, Aabloo A,
  Djurabekova F and Zadin V 2018
  \href{https://doi.org/https://doi.org/10.1088/1361-6528/aa9a1b}{{\em
  Nanotechnology\/} {\bf 29} 015704} (\textit{Preprint}
  \href{https://arxiv.org/abs/1709.09104}{{\tt 1709.09104}})
  \urlprefix\url{https://doi.org/10.1088/1361-6528/aa9a1b}

\bibitem{jansson2016long}
Jansson V, Baibuz E and Djurabekova F 2016 {\em Nanotechnology\/} {\bf 27}
  265708 (\textit{Preprint} \href{https://arxiv.org/abs/1508.06870}{{\tt
  1508.06870}}) \urlprefix\url{https://doi.org/10.1088/0957-4484/27/26/265708}

\bibitem{kimocs}
 2014 {Kimocs --- a Kinetic Monte Carlo simulation code for surfaces}
  {Available under the terms of the GNU General Public License.}
  \urlprefix\url{https://gitlab.com/vjansson/Kimocs}

\bibitem{baibuz2018migration}
Baibuz E, Vigonski S, Lahtinen J, Zhao J, Jansson V, Zadin V and Djurabekova F
  2018
  \href{https://doi.org/https://doi.org/10.1016/j.commatsci.2017.12.054}{{\em
  Computational Materials Science\/} {\bf 146}(C) 287--302}
  \urlprefix\url{https://doi.org/10.1016/j.commatsci.2017.12.054}

\bibitem{lahtinen2018artificial}
Lahtinen J, Jansson V, Vigonski S, Baibuz E, Domingos R, Zadin V and
  Djurabekova F 2018 {\em arXiv:1806.02976 [physics.comp-ph]\/} Submitted for
  publication \urlprefix\url{https://arxiv.org/abs/1806.02976}

\bibitem{zhao2016formation}
Zhao J, Baibuz E, Vernieres J, Grammatikopoulos P, Jansson V, Nagel M,
  Steinhauer S, Sowwan M, Kuronen A, Nordlund K {\em et~al.\/} 2016 {\em ACS
  nano\/} {\bf 10} 4684--94
  \urlprefix\url{https://doi.org/10.1021/acsnano.6b01024}

\bibitem{antczak2004long}
Antczak G and Ehrlich G 2004 {\em Physical Review Letters\/} {\bf 92} 166105

\bibitem{mills1994quantum}
Mills G and J{\'o}nsson H 1994 {\em Physical Review Letters\/} {\bf 72} 1124

\bibitem{mills1995reversible}
Mills G, J{\'o}nsson H and Schenter G~K 1995 {\em Surface Science\/} {\bf 324}
  305--37

\bibitem{lammps}
Plimpton S 1995 {\em Journal of computational physics\/} {\bf 117} 1--19

\bibitem{marinica2013interatomic}
Marinica M~C, Ventelon L, Gilbert M, Proville L, Dudarev S, Marian J, Bencteux
  G and Willaime F 2013 {\em Journal of Physics: Condensed Matter\/} {\bf 25}
  395502

\bibitem{sand2016non}
Sand A, Dequeker J, Becquart C, Domain C and Nordlund K 2016 {\em Journal of
  Nuclear Materials\/} {\bf 470} 119--27

\bibitem{derlet2007multiscale}
Derlet P~M, Nguyen-Manh D and Dudarev S 2007 {\em Physical Review B\/} {\bf 76}
  054107

\bibitem{debe1977space}
Debe M and King D~A 1977 {\em Physical Review Letters\/} {\bf 39} 708

\bibitem{heinola2010first}
Heinola K and Ahlgren T 2010
  \href{https://doi.org/10.1103/PhysRevB.81.073409}{{\em Phys. Rev. B\/} {\bf
  81}(7) 073409}

\bibitem{olewicz2014coexistence}
Olewicz T, Antczak G, Jurczyszyn L, Lyding J~W and Ehrlich G 2014
  \href{https://doi.org/10.1103/PhysRevB.89.235408}{{\em Phys. Rev. B\/} {\bf
  89}(23) 235408}

\bibitem{chen2013biaxial}
Chen Z and Ghoniem N 2013
  \href{https://doi.org/10.1103/PhysRevB.88.035415}{{\em Phys. Rev. B\/} {\bf
  88}(3) 035415}

\bibitem{ovito}
Stukowski A 2010 {\em Modelling and Simulation in Materials Science and
  Engineering\/} {\bf 18} 015012 \urlprefix\url{http://ovito.org/}

\bibitem{liu1991eam}
Liu C, Cohen J, Adams J and Voter A 1991 {\em Surface science\/} {\bf 253}
  334--44

\bibitem{bukonte2014modelling}
Bukonte L, Ahlgren T and Heinola K 2014 {\em Journal of Applied Physics\/} {\bf
  115} 123504

\bibitem{mundy1987vacancy}
Mundy J, Ockers S and Smedskjaer L 1987 {\em Philosophical Magazine A\/} {\bf
  56} 851--60

\bibitem{ase}
Larsen A~H, Mortensen J~J, Blomqvist J, Castelli I~E, Christensen R, Du{\l}ak
  M, Friis J, Groves M~N, Hammer B, Hargus C {\em et~al.\/} 2017 {\em Journal
  of Physics: Condensed Matter\/} {\bf 29} 273002

\bibitem{vitos1998surface}
Vitos L, Ruban A, Skriver H~L and Kollar J 1998 {\em Surface Science\/} {\bf
  411} 186--202

\end{thebibliography}

\end{document}